\documentclass[11pt,sort&compress]{elsarticle}
\input{preamble.tex}

\begin{document}

\hypersetup{
  linkcolor=darkrust,
  citecolor=seagreen,
  urlcolor=darkrust,
  pdfauthor=author,
}

\begin{frontmatter}

\title{{\large\bfseries Coupled Rayleigh--Taylor and Faraday instabilities \\ in vertically vibrated cylindrical containers}}

\author[gtcse]{Tianyi~Chu\corref{cor1}}
\ead{tchu72@gatech.edu}
\author[gtcse]{Benjamin~Wilfong}
\author[sandia]{Timothy~Koehler}
\author[sandia]{Ryan~M.~McMullen}
\author[gtcse,gtae,gtme]{Spencer~H.~Bryngelson}

\cortext[cor1]{Contact author.}

\address[gtcse]{School of Computational Science \& Engineering, Georgia Institute of Technology, Atlanta, GA 30332, USA \vspace{-0.125cm}}
\address[gtae]{Daniel Guggenheim School of Aerospace Engineering, Georgia Institute of Technology, Atlanta, GA 30332, USA \vspace{-0.125cm}}
\address[gtme]{George~W.~Woodruff School of Mechanical Engineering, Georgia Institute of Technology, Atlanta, GA 30332, USA \vspace{-0.125cm}}
\address[sandia]{Engineering Sciences Center, Sandia National Laboratories, P.O.~Box 5800, Albuquerque, NM 87185, USA}

\date{}

\end{frontmatter}

\begin{abstract}
Interfacial instabilities govern the mixing in confined multiphase flows.
Yet, the two mechanisms that drive them are usually studied independently: the pressure-gradient-driven Rayleigh--Taylor (RT) instability, which amplifies long-wavelength modes, and the parametrically forced Faraday instability, which selects shorter-wavelength harmonic or subharmonic modes.
When an adverse density contrast and vertical vibration act together, the two compete, and neither alone describes the response.
We use Floquet analysis to characterize the onset, growth, modal structure, and velocity fields of Faraday--RT waves in a vertically vibrated cylinder, resolved by azimuthal wavenumber, radial (Bessel) mode, and Floquet harmonic.
The formulation recovers the classical RT and Faraday limits and reproduces the instability onset at the frequencies measured experimentally.
For a free-sliding interface, increasing the vibration amplitude shifts the dominant instability mechanism from RT growth to subharmonic and then harmonic Faraday responses.
Lateral confinement can also stabilize individual RT modes, which is not possible in an unbounded domain, although other Faraday modes may remain unstable.
Pinning the contact line couples radial modes that otherwise evolve independently, allowing the unstable mode to be a superposition of RT-unstable and Faraday-stable components.
This superposition alters the instability mechanism, producing a richer radial pattern.
Reconstruction of the unstable modes shows the (linear) velocity fields that imaging cannot access and demonstrates how the instabilities can change the flow more broadly.
\end{abstract}

\section{Introduction}

In confined multiphase flows, interface deformation can be driven by the interplay among fluid-property contrasts, external forcing, geometry, and boundary conditions.
The growth of interfacial perturbations can lead to interface breakup, entrainment, mixing, and bubble injection, as observed, for example, in vibration-forced drop atomization~\citep{james2003vibration,vukasinovic2007dynamics}.
In many engineering and physical settings, these processes occur in laterally confined geometries, such as liquid sloshing in fuel tanks~\citep{ibrahim2005liquid} and interfacial dynamics in inertial confinement fusion configurations~\citep{craxton2015direct, betti2016inertial}.
Contact-line constraints are likewise common in brimful containers, sessile and pendant droplets~\citep{mohammad2022review}, coating and printing processes~\citep{snoeijer2013moving}, and droplet-based microfluidic systems~\citep{holmes2015transporting}.
Pinning at a solid boundary, such as the rim of a cylindrical container, can alter the admissible interfacial modes and shift the corresponding instability thresholds, making interfacial dynamics more challenging to predict.
Understanding the underlying instability mechanisms in cylindrical coordinates is therefore fundamental for predicting and controlling such confined interfacial flows.

Linear stability analysis identifies two canonical hydrodynamic instability mechanisms in vibrated containers.
When the vibration amplitude exceeds a critical threshold, the flat interface destabilizes via parametric resonance, producing standing surface waves known as Faraday waves~\citep{faraday1831xvii}.
Separately, an adverse density contrast under gravity can generate a pressure-gradient-driven instability, leading to the Rayleigh--Taylor (RT) instability~\citep{rayleigh1882investigation, taylor1950instability}.
Readers are referred to reviews of these two instabilities by \citet{miles1990parametrically} for Faraday waves and by \citet{kull1991theory} for RT instability.
Both mechanisms are central to a wide range of natural processes and engineering applications, and each has been studied extensively.
Particular attention has been given to confined cylindrical geometries because of their prevalence and their role as canonical configurations for finite-domain interfacial dynamics.
In such geometry, finite-domain effects, azimuthal mode selection, and contact-line conditions can substantially modify the interfacial response.
Coupled Faraday--RT behavior arises most directly in vertically vibrated, adversely stratified liquid layers, including vibration-stabilized levitating liquids~\citep{apffel2020floating} and liquid films~\citep{pototsky2016faraday,sterman2017rayleigh,gao2026marangoni}.
More broadly, insights into this multi-instability regime may inform the design and control of vibrated multiphase systems, vibration-assisted coating and printing processes, and fluid management in partially filled propellant tanks subjected to simultaneous acceleration and structural vibration.
Motivated by these applications, this work examines how cylindrical confinement and contact-line conditions govern the onset and coupling between Faraday and RT instabilities.

In laterally unbounded domains, Faraday instabilities are commonly studied by decomposing interfacial disturbances into Fourier modes.
Early linear stability analyses described the free-surface dynamics of an ideal fluid in terms of Mathieu equations~\citep{benjamin1954stability}.
These analyses were subsequently extended to damped Mathieu equations to account for weak viscous damping~\citep{landau1976mechanics}, and later to Floquet analyses of viscous fluid--fluid interfaces with capillary effects~\citep{kumar1994parametric,kumar1996}.
Experimentally, \citet{edwards1994patterns} demonstrated that, in contrast to the parallel-line patterns observed under single-frequency forcing, multi-frequency forcing can excite more complex $N$-fold rotationally symmetric patterns, such as squares and hexagons.
These observations indicate that multiple resonant Faraday waves can be excited simultaneously.
In parallel, secondary instabilities beyond the linear regime have been investigated~\citep{zhang1995secondary,daudet1995secondary}, arising from nonlinear interactions among standing Faraday waves.
Weakly nonlinear theories were subsequently developed to predict the stability of such patterns in the vibration frequency--amplitude parameter space~\citep{zhang1996square,chen1997pattern,chen1999pattern}, with experimental validation by \citet{westra2003patterns}.
More recently, \citet{panda2025marangoni} numerically showed that increasing Marangoni stresses can drive transitions from squares to asymmetric Faraday patterns.
\citet{castillo2025mixing} combined experiments and numerical simulations to examine parametrically induced mixing across the interface between two miscible fluids, showing that the dominant modal structures remain reminiscent of Fourier modes.

For Faraday waves in confined container geometries, extensive attention has been given to cylindrical configurations due to their prevalence in engineering applications.
Under free-sliding boundary conditions, Hamiltonian formulations have been used to analyze surface-wave damping and nonlinear dynamics~\citep{miles1967surface,miles1976nonlinear,miles1984nonlinear}.
Linear Floquet stability predictions have also been validated against measurements of Faraday thresholds at low forcing frequencies~\citep{batson2013faraday}, and against observed Bessel-mode pattern formation in forced cylindrical configurations~\citep{dinesh2023pattern}.
Brimful cylinders have attracted particular interest because they provide a controlled realization of a pinned circular contact line.
In this configuration, the rim constraint alters the admissible modal structure and the corresponding onset conditions.
By constructing appropriate superpositions of radial Bessel modes to satisfy this boundary condition, \citet{henderson1994surface} used linear theory to predict the natural frequencies and damping ratios of surface modes in a brimful cylinder, obtaining good agreement with experiments.
\citet{kidambi2013inviscid} formulated a set of coupled Mathieu equations to obtain the inviscid combination resonance tongues (CRTs) associated with the admissible radial structures.
More recently, \citet{shao2021surface} conducted mechanical-vibration experiments in brimful cylinders and showed that inviscid theory can accurately predict the observed surface patterns and natural frequencies at a water--air interface.
\citet{zhang2023pattern} further demonstrated experimentally that the pinned-contact-line constraint can induce not only linear coupling among radial modes with the same azimuthal wavenumber, but also among harmonic azimuthal components.
Subsequent experimental studies have examined these modal interactions by tracking the pattern evolution over each vibration period~\citep{gregory2024tracking,zhang2025transitional}.
In parallel, \citet{bongarzone2022subharmonic} used weakly nonlinear analysis to examine the effect of the contact angle on the onset of subharmonic Faraday patterns and validated the results with numerical simulations.
Together, these studies establish a strong foundation for Faraday-wave pattern formation in cylindrical containers, with emphasis on regimes where parametric resonance is the dominant instability mechanism and the dynamics remain close to neutral stability.

At larger vibration amplitudes, the interface can undergo RT growth during part of the forcing cycle, when the instantaneous effective gravity points from the denser phase toward the lighter phase~\citep{kumar2000mechanism,wright2000numerical}.
This periodic alternation between RT-stabilizing and RT-destabilizing phases 
makes a conventional exponential normal-mode description alone insufficient for capturing the transient interfacial dynamics, motivating the use of Floquet analysis over the full forcing period.
In the absence of time-periodic forcing, the analysis of RT instability originated from studies of small-amplitude interfacial waves in density-contrast fluids.
The RT theory has progressed from inviscid~\citep{rayleigh1882investigation,taylor1950instability} to viscous extensions~\citep{harrison1908influence} and subsequently to models incorporating surface tension~\citep{bellman1954effects}.
Compared with the resonance-selected patterns associated with Faraday waves, the RT mechanism preferentially amplifies longer-wavelength interfacial modes.
The asymptotic limits of large and small density and viscosity contrasts were investigated by \citet{plesset1974viscous}.
Much of the classical theory of RT instability was comprehensively summarized by \citet{chandrasekhar1961hydrodynamic}, with subsequent developments reviewed by \citet{sharp1983overview} and \citet{kull1991theory}.
By balancing the relevant forces in non-ideal fluids, \citet{piriz2006} derived a simple yet accurate analytical model for RT instability.
RT instability has also been examined in cylindrical confinement, where the lateral boundary and the azimuthal mode structure play essential roles in the instability mechanism.
For stratified fluids in a vertical cylinder, \citet{batchelor1993instability} showed that the rigid circular boundary regularizes the neutral-stability problem by increasing the critical Rayleigh number from zero in the unbounded limit to a finite value.
Related density-driven instabilities in capillary tubes have been investigated for miscible interfaces using linear stability analysis~\citep{vanaparthy2003density,payr2005influence}.
\citet{sweeney2013rayleigh} combined linear stability theory with experimental observations of long-time fingering states to show that fluid-property contrasts affect the preferred azimuthal mode of RT instability.

The pinned-contact-line constraint has received comparatively limited attention in studies of RT instability because large interfacial deformations make it difficult to maintain the fixed-boundary condition in Newtonian fluids.
In soft or structured media, however, additional restoring stresses can help suppress the boundary displacement.
For example, \citet{zheng2019rayleigh} experimentally showed that RT waves in an elastic soft gel can remain pinned at the cylinder rim.
Similar suppression of boundary displacement can also arise in the presence of a foam~\citep{bret2026linear}.
In the absence of these additional stresses, alternative stabilization mechanisms are therefore needed to study RT instability under a pinned-contact-line constraint.

Vertical vibration can induce or amplify Faraday instabilities, but it can also dynamically suppress RT growth.
Depending on the fluid properties and container geometry, there may exist a range of forcing amplitudes and frequencies over which an otherwise RT-unstable interface can be dynamically stabilized \textit{without} exciting standing Faraday waves.
Early studies of such dynamical stabilization can be traced to \citet{wolf1969dynamic,wolf1970dynamic}, who experimentally demonstrated that vertical vibration can stabilize an RT-unstable interface in a cylindrical container when viscous effects are sufficiently strong.
The theoretical analysis of \citet{Troyon1971} subsequently supported this conclusion by deriving stability criteria that relate the required forcing amplitude to the container size and fluid properties, and further noted that finite surface tension is required for complete stabilization.
This role of surface tension was later supported analytically by \citet{piriz2010dynamic}.
Related stabilization mechanisms in cylindrical geometries have also been examined theoretically in the thin-film limit
~\citep{lapuerta2001control, haimovich2010nonlinear}.
More recently, \citet{liang2025experimental} conducted an experimental study of a water--air interface and showed that the onset and growth of cylindrical RT modes depend on the azimuthal wavenumber.

\begin{figure}[ht!]
  \centering
  \includegraphics[scale=1]{ 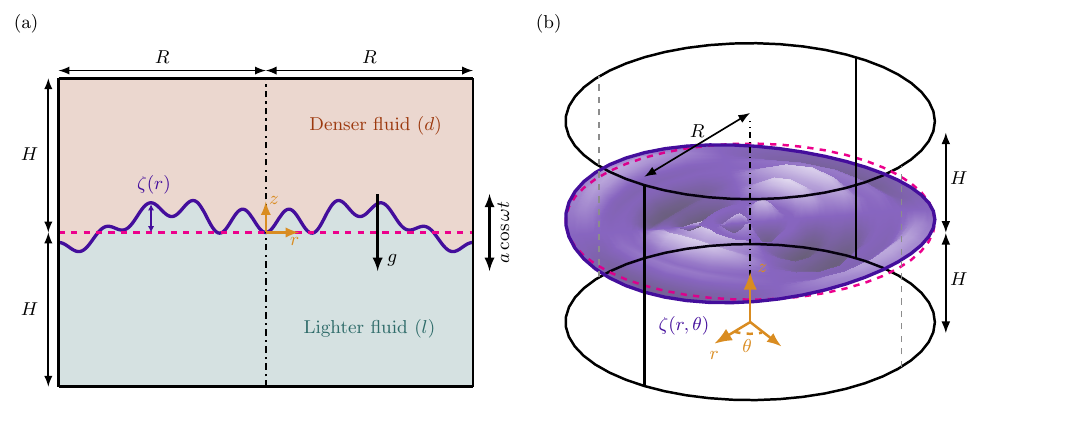}
  \caption{
    Schematic of coexisting long-wavelength RT and shorter-wavelength Faraday disturbances at a two-fluid interface in a confined cylinder: (a) 2-D cross-section and (b) 3-D interface.
    The quadrupolar component ($m=2$) is shown as an example, with the unperturbed contact line at the cylinder rim highlighted.
  }
  \label{f:overview}
\end{figure}

Despite the extensive literature on Faraday and RT waves separately, comparatively few studies have addressed regimes in which the two mechanisms \textit{coexist} and jointly deform the interface.
\Cref{f:overview} shows a schematic of the two-fluid interface in a confined cylinder, where both long-wavelength RT and short-wavelength Faraday waves coexist. 
Our prior work showed competition between long-wavelength RT modes and shorter-wavelength Faraday modes in an unbounded two-dimensional (2-D) domain and validated Floquet-predicted interfacial dynamics using direct simulations~\citep{chu2025competing}.
This work extends the analysis to a Floquet-based 3-D configuration with different boundary conditions, including confinement.
A distinct set of boundary conditions and higher-dimensional geometry demonstrates the rich, otherwise unobserved behavior the 2-D problem exhibited, including azimuthal and radial ones.
At the same time, the lateral wall discretizes the admissible modes and enables their coupling.
The additional azimuthal degree of freedom generates families of non-axisymmetric interfacial patterns, each governed by a distinct instability mechanism.
We characterize the onset, modal structure, and flow response of mixed Faraday–RT instabilities for both free-sliding and pinned contact lines.
We show that confinement can stabilize an RT mode over a finite range of forcing amplitudes, which is not the case in unbounded domains.
In the unbounded case, a set of long-wavelength modes will always remain unstable.
Pinning the contact line couples radial modes that otherwise evolve independently.
Thus, the unstable mode is a superposition of RT-unstable and Faraday-stable components.
As such, the mechanism of this onset also changes, more than shifting a stability threshold.
We also reconstruct the linear velocity fields associated with each response, unavailable to experimental imaging.

This paper is organized as follows.
\Cref{background} describes the governing equations, and in \cref{sec: LFST} we perform a linear Floquet stability analysis to characterize the interfacial dynamics and the associated flow-field response.
Owing to the rotational homogeneity and the time-periodic forcing, we decompose the spatiotemporal interfacial response into individual azimuthal wavenumbers and Floquet harmonics, enabling the separate investigations of RT and Faraday modes.
In \cref{results}, we explore the vibration frequency--amplitude parameter space to identify how the dominant instability mechanism transitions across regimes and to show the additional linear coupling between RT and Faraday modes induced by the pinned-contact-line constraint.
\Cref{discussion} discusses the limitations and outlook of the analysis and summarizes the main findings on the coexistence and competition between RT and Faraday instabilities.

\section{Model equations}\label{background}

Consider the interface between two immiscible and incompressible fluids: a denser and lighter fluid, denoted by superscripts $\left(\cdot\right)^{(d)}$ and $\left(\cdot\right)^{(l)}$. 
The two fluids occupy a cylindrical vessel of height $2H$ and radius $R$, with the interface at the mid-height $H$.
The container is subject to constant gravitational acceleration and an oscillatory vertical acceleration
\begin{equation}
    \tilde{g}(t) = g_{\text{sgn}}g + a \cos (\omega t+\varphi_0),
\end{equation}
where $a$ is the oscillatory amplitude, $\omega$ is the corresponding frequency, and $\varphi_0$ is the initial phase.
Positive gravity, $g_{\text{sgn}}=1$, applies when the denser fluid is at the bottom domain, and negative gravity, $g_{\text{sgn}}=-1$, applies when the denser fluid is on top.
\Cref{f:overview} shows the two-fluid interface subjected to a harmonic vibration.
Within each fluid layer $j$, the motion of the fluid is governed by the incompressible Navier--Stokes equations,
    \begin{align}\label{eqn:NS}
    \rho^{(j)} \left[\partial_t +\left(\vb*{u}^{(j)}\cdot \vb*{\nabla}\right) \right]\vb*{u}^{(j)}=-\grad p^{(j)}-\rho^{(j)} \tilde{g}\vb*{e}_z +\mu^{(j)} \grad^2\vb*{u}^{(j)},\quad 
    \nabla\cdot \vb*{u}^{(j)} =0.
\end{align}
To analyze the interfacial dynamics, we nondimensionalize the governing equations using a characteristic length $H$ and gravitational acceleration $g$.
The gravitational time follows as $t_c=\sqrt{H/g}$.
Density $\rho$, viscosity $\mu$, and surface tension $\sigma$, are scaled using appropriate reference values.
\Cref{dimensionless_quantity} summarizes the resulting dimensionless variables, denoted by $(\cdot)^*$.

\begin{table}[ht!]
  \centering
  \caption{Dimensionless quantities.}\label{dimensionless_quantity}
  {\setlength{\tabcolsep}{5pt}\small
  \begin{tabular}{l l}
     Dimensionless quantity  & Quantity name       \\ \midrule
    $\xi=H/R$  & Cylinder aspect ratio\\
      $\At=(\rho^{(d)}-\rho^{(l)})/(\rho^{(d)}+\rho^{(l)})$ & Atwood number \\
     $\eta = \mu^{(l)}/\mu^{(d)}$ & Viscosity ratio \\
     $C= \nu^{(d)}/\sqrt{gH^3}$ & Viscous-to-gravitational force ratio\\
     $\mathrm{Bd} = \rho^{(d)} gH^2/\sigma^{(d)}$ & Bond number\\
       $a^*=a/g$            & Acceleration  \\
     $\zeta^*=\zeta/H$             & Interface displacement   \\
        $t^*=t/t_c$              & Time\\
       $\gamma^*=\gamma t_c$            & Growth rate  \\
        $\omega^*=\omega t_c$ &   Oscillatory frequency  \\
      $[u_r, u_\theta,w]^* = [u_r, u_\theta,w] t_c/ H$          & Velocity components
  \end{tabular}
  }
\end{table}

Equation~\cref{eqn:NS} is nondimensionalized as
    \begin{align}\label{eqn:full NS}
    \left[\partial_{t^*} +\left({\vb*{u}^{*^{(j)}}}\cdot \vb*{\nabla}^*\right)\right] \vb*{u}^{*^{(j)}}&=-\grad^* p^{*^{(j)}}- \tilde{g}^*\vb*{e}_z +C^{(j)} {\boldsymbol{\Delta}^*}\vb*{u}^{*^{(j)}},\quad 
    \grad^*\cdot \vb*{u}^{*^{(j)}} =0,
\end{align}
where $C^{(d)}=C$ and $C^{(l)} = \eta C(1+\At)/(1-\At)$.
We decompose the flow state around the equilibrium-state solution of the Navier--Stokes equations~\cref{eqn:full NS} as
\begin{align}
  \vb*{u}^{*^{(j)}}=\vb*{U}^{*^{(j)}}+ {\vb*{u}'^{*^{(j)}}},\quad  p^{*^{(j)}}={P}^{*^{(j)}}+  {p'^{*^{(j)}}},
\end{align}
where $(\vb*{U}^{*^{(j)}}, \grad^* P^{*^{(j)}}) = (0,- \tilde{g}^*\vb*{e}_z )$ and $(\,\cdot\,)'$ denotes the small fluctuating components.
The linearized governing equations for the fluctuations are
\begin{align}\label{eqn:NS fluctuating}
     \partial_{t^*} {\vb*{u}'^{*^{(j)}}} =-\grad^* {p'^{*^{(j)}}} +C^{(j)} {\boldsymbol{\Delta}^*}{\vb*{u}'^{*^{(j)}}}, \quad 
    \grad^* \cdot {\vb*{u}'^{*^{(j)}}} =0.
\end{align}
Applying $\vb*{e}_z\cdot \grad^*\times\grad^*\times$ to equation~\cref{eqn:NS fluctuating} eliminates the horizontal velocity components, yielding
\begin{align}\label{eqn:NS fluctuating_w}
    \left(\partial_{t^*}-C^{(j)}{\boldsymbol{\Delta}^*}\right){\boldsymbol{\Delta}^*} {w'^{*^{(j)}}}=0,
\end{align}
Continuity of ${w'^{*^{(j)}}}$ and the tangential stress at the interface $z^*=\zeta^*$ gives
\begin{align}\label{eqn:interface_cont}
    \delta \left[{w'^{*^{(j)}}},\, \partial_{z^*}\left( {w'^{*^{(j)}}}\right),\, C^{(j)}\left(\boldsymbol{\Delta}^*-2\partial_{z^*z^*}\right){w'^{*^{(j)}}}\right]=0,
\end{align}
where $\delta$ represents the interface jump. 
The linearized kinematic boundary condition gives
\begin{align}\label{eqn:kinematic}
    \partial_{t^*} {\zeta}^* &={w}^*\vert_{z^*=0}.
\end{align}
The jump condition for pressure across the interface,
\begin{align}
    \label{eqn:pressure jump}
    \delta p^* &= 2(C^{(l)}-C^{(d)})
    \left(\partial_{z^*}{w}^*\right)\vert_{z^*=0}-\frac{2\mathrm{At}}{1+\mathrm{At}}\tilde{g}^*\zeta^*+\frac{1}{\mathrm{Bd}}\left(\boldsymbol{\Delta}^*-\partial_{z^*z^*}\right) \zeta^*,
\end{align}
explicitly represents the effect of vibration.
Together, equations~\Crefrange{eqn:NS fluctuating_w}{eqn:pressure jump} govern the linearized interfacial dynamics. 
At later times, as nonlinear effects become significant, the interface may evolve into a complex profile and undergo breakup.
Here, we focus on the regime in which the interface profile $\zeta^*$ remains well-defined and single-valued.
The superscript $(\cdot)^*$ denoting dimensionless variables is omitted in \cref{sec: LFST}.
All variables are nondimensional unless otherwise specified.

\section{Linear Floquet stability analysis in cylindrical coordinates}\label{sec: LFST}

We use the rotational symmetry of the problem to expand both the velocity field and the interface displacement in an azimuthal Fourier series.
Because the instability is hypothesized to be driven mainly by time-periodic variations in gravity, we model the dynamics using Floquet analysis~\citep{floquet1883equations}.
The vertical displacement of the interface at each azimuthal wavenumber is expected to exhibit exponential growth or decay with superimposed oscillatory fluctuations.
The spatiotemporal decomposition of the vertical velocity perturbation and interface displacement is, thus,
\begin{align}
    \left[{{w'}^{(j)}}(\vb*{x},z, t),\, \zeta(\vb*{x}, t)\right] = \sum_{m=-\infty}^{\infty}\sum_{i=1}^{\infty}
    \underbrace{\left(\sum_{n=-\infty}^{\infty}
     \left[{\hat{w}}_{m,n,i}^{(j)}(r,z),\, \hat{\zeta}_{m,n,i}(r)\right]\rme^{\rmi n \omega (t+t_0)}\right)}_{\text {Periodic component}}\underbrace{\rme^{\gamma_{m,\cdot,i} t}}_{\text{Modal}}\underbrace{\rme^{\rmi m \theta}}_{\text{Azimuthal}}, \label{eqn:decomposition_w} 
\end{align}
where $\gamma_{m,\cdot,i}$ denotes the complex Floquet exponent of the
$i$th radial mode at azimuthal wavenumber $m$, $t_0$ is the time shift associated with the initial phase $\varphi_0$, and $n$ is the integer index of the harmonics \citep{kumar1994parametric,kumar1996}.
A positive real part of the Floquet exponent indicates instability, leading to the growth and eventual interfacial breakup.
Equation~\cref{eqn:decomposition_w} decomposes the waveforms into different azimuthal wavenumbers and oscillation harmonics.
Substituting this decomposition into the governing equation~\cref{eqn:NS fluctuating_w} for a given triplet $(m,n,i)$ gives
\begin{align}
  \label{eqn:Governing cylindrical}
  \left[\gamma_{m,n,i}-C^{(j)}\boldsymbol{\Delta}_m\right]\boldsymbol{\Delta}_m {\hat{w}_{m,n,i}}^{(j)}=0,
\end{align}
where
\begin{align}
  \boldsymbol{\Delta}_m \equiv 
    \boldsymbol{\Delta}_m^{\text{H}}+\partial_{zz}\equiv \partial_{rr} + 
        \frac{1}{r}\partial_{r}-\frac{m^2}{r^2}+\partial_{zz}
\end{align}
is the Laplace operator associated with the $m$th azimuthal component, and $\gamma_{m,n,i}\equiv\gamma_{m,\cdot,i}+\rmi n\omega$ is the complex growth rate of the $n$th temporal harmonic.
The solution to equation~\cref{eqn:Governing cylindrical} is
\begin{align}
  \hat{w}_{m,n,i}^{(j)} (r,z)= \frac{J_{|m|}(\beta_{m,n,i} r)}{J_{|m|}(\beta_{m,n,i}/\xi)}  {Z}_{m,n,i}^{(j)}(z),
\end{align}
where $\beta_{m,n,i}$ is the $i$th radial wavenumber satisfying
\begin{align}
    J'_{|m|}(\beta_{m,n,i}/\xi)=0. 
\end{align}
This radial wavenumber is independent of the fluid phase $j$ and the Floquet harmonic index $n$.
Henceforth, we omit the dependence of $\beta$ on $n$ for simplicity and use the notation $\beta_{m,\cdot,i}$ for consistency with the other modal quantities.
The axial structure associated with each Floquet harmonic is then expressed in the general form
\begin{align}\label{eqn:w_genearal_app}
 Z^{(j)}_{m,n,i}(z) 
  &= \hat{\zeta}_{m,n,i} 
    \left(
    a^{(j)}_{m,n,i}  \mathrm{e}^{\beta_{m,\cdot,i} z} +
    b^{(j)}_{m,n,i} \mathrm{e}^{-\beta_{m,\cdot,i} z} +
    c^{(j)}_{m,n,i} \mathrm{e}^{ \beta_{m,\cdot,i} q_{m,n,i}^{(j)} z} +
    d^{(j)}_{m,n,i} \mathrm{e}^{-\beta_{m,\cdot,i} q_{m,n,i}^{(j)} z}
    \right),
\end{align}
where $q_{m,n,i}^{(j)}\equiv\sqrt{1+\gamma_{m,n,i}/(C^{(j)}\beta^2_{m,\cdot,i})}$.
The modal coefficients are determined by enforcing the boundary and interfacial conditions. 
\Cref{sec: Floquet_App} provides the full derivation.

To relate the adjacent Floquet harmonics,  we rewrite the pressure-jump condition in equation~\cref{eqn:pressure jump} in modal form as
\begin{align}\label{eqn:pressure jump_coeffs}
& a \left(\hat{\zeta}_{m,n-1,\cdot}+\hat{\zeta}_{m,n+1,\cdot}\right)
- \left(2g_{\text{sgn}} -\frac{\kappa}{\mathrm{Bd}} \boldsymbol{\Delta}_m^{\mathrm{H}}\right) \hat{\zeta}_{m,n,\cdot} \nonumber \\
= &
    \sum_i    \left\{ \left[ \frac{\gamma_{m,n,i}}{\beta_{m,\cdot,i}^2}+ 3\kappa C(1-\eta)  \right]\partial_z \hat{w}_{m,n,i}^{(d)} -\frac{\kappa C}{\beta_{m,\cdot,i}^2}\left[\partial_{zzz}\hat{w}_{m,n,i}^{(d)}-\eta \partial_{zzz}\hat{w}_{m,n,i}^{(l)}\right]\right\} \frac{J_{|m|}(\beta_{m,\cdot,i} r)}{J_{|m|}(\beta_{m,\cdot,i}/\xi)}
\end{align}
where $\kappa \equiv (1+\At)/(\At)$ for brevity.
The total interfacial displacement at the $n$th Floquet harmonic can then be expressed as a superposition of particular and homogeneous contributions,
\begin{align}\label{eqn:total displacement}
  \hat{\zeta}_{m,n,\cdot} = 
    -\hat{\zeta}_{m,n,k}^{(\mathrm{H})} \frac{I_{|m|}( r/l_{m,n,k})}{I_{|m|}
        (1/(\xi l_{m,n,k}))}+\sum_{i}\hat{\zeta}_{m,n,i}^{(\mathrm{P})}
        \frac{J_{|m|}(\beta_{m,\cdot,i} r)}{J_{|m|}(\beta_{m,\cdot,i}/\xi)}.
\end{align}
Here, $I_{|m|}(\cdot)$ are the modified Bessel functions of the first kind of order $|m|$, and $l_{m,n,k}$ is
the corresponding characteristic capillary length.
The particular component has the same radial dependence as the velocity perturbations, $J_{|m|}(\beta_{m,\cdot,i}r)$.
The homogeneous solution arises from the surface tension effect~\cref{eqn:pressure jump_coeffs},
\begin{align}\label{eqn:interface_homo}
  \left(2g_{\text{sgn}} - \frac{\kappa}{\mathrm{Bd}\,l_{m,n,k}^2} \right) \hat{\zeta}_{m,n,k}^{(\mathrm{H})} =
    a \left(
\hat{\zeta}_{m,n+1,k}^{(\mathrm{H})}+\hat{\zeta}_{m,n-1,k}^{(\mathrm{H})}
    \right).
\end{align}
The coupling between adjacent Floquet harmonics may be written as
\begin{align}\label{eqn harmonic_relation_homo}
\hat{\zeta}_{m,n+1,k}^{(\mathrm{H})}+\hat{\zeta}_{m,n-1,k}^{(\mathrm{H})}=2\cos{\varphi_{m,n,k}}\,
    \hat{\zeta}_{m,n,k}^{(\mathrm{H})},
\end{align}
and the corresponding capillary length scales are
\begin{align}\label{eqn: capillary length}
    l_{m,n,k}^2= {\frac{\kappa}{2\mathrm{Bd}\left(g_{\text{sgn}}-a\cos{\varphi_{m,n,k}}\right)}}.
\end{align}
In the absence of vibration, $a=0$, and for the Faraday-only configuration $g_{\text{sgn}}=1$, this reduces to the static capillary length $l_{m,n,k}=\sqrt{{\kappa/2\mathrm{Bd}}}$.
The $I_{|m|}$-type homogeneous solution exists only if
$g_{\text{sgn}}+a>0$, such that $l_{m,n,k}^{2}>0$.
When this quantity is negative, the capillary length becomes imaginary, and the homogeneous solution reduces to an oscillatory $J_{|m|}$-type radial dependence.
Here $\varphi_{m,n,k}$ is the Floquet phase of the homogeneous (capillary) subsystem.
As with the harmonic coupling in equation~\cref{eqn harmonic_relation_homo}, the homogeneous relation alone does not fix it, leaving an admissible band $\cos\varphi_{m,n,k}\in[-1,1]$.
We close the problem by retaining the band edge $\cos\varphi_{m,n,k}=-1$, which yields the smallest positive capillary length,
$l^2_{m,\cdot,k}=\kappa/[2\mathrm{Bd}(g_{\text{sgn}}+a)]$,
and thus the largest vibrational contribution to the capillary restoring stress.
The onset thresholds reported below, therefore, correspond to an isolation of the strongest stabilizing effect of the vibration.

\subsection{Free-sliding interface}

We first consider the free-sliding configuration, in which no additional constraint is imposed on the interfacial displacement $\zeta$.
In this case, the homogeneous contribution vanishes, $\hat{\zeta}_{m,n,k}^{(\mathrm{H})}=0$, allowing each radial mode to evolve independently.
The governing equation for each azimuthal--radial mode family can be compactly expressed as a generalized eigenvalue problem
\begin{align}
    \vb*{A}_{m,\cdot,i}(\gamma_{m,\cdot,i};\beta_{m,\cdot,i})\hat{\vb*{\zeta}}_{m,\cdot,i}=a\vb*{B} \hat{\vb*{\zeta}}_{m,\cdot,i},
\end{align}
with details provided in \cref{sec: Floquet_App}.
The most unstable Floquet exponent is obtained from
\begin{align}
    \gamma_{m,\cdot,i}^{(U)}= \argmax_{\det{\vb*{A}_{m,\cdot,i}-a\vb*{B}} = 0}  \mathrm{Re}\{\gamma_{m,\cdot,i}\},
    \label{eqn:Det}
\end{align}
which selects the root with the largest real part and gives the leading linear growth rate.
Unlike Floquet exponents obtained from the eigenvalues of the monodromy matrix, the present formulation selects a single representative leading exponent for each radial mode.
The periodic harmonic amplitudes, $\hat{\vb*{\zeta}}_{m,\cdot,i}$, which describe the oscillatory modulation about the exponential growth or decay, are then obtained from the null space of $\vb*{A}_{m,\cdot,i}-a\vb*{B}$.
Equation~\cref{eqn:Det} provides a generalized analysis of the hydrodynamic instability in fluid mixing, accounting for finite fluid-layer depth, viscosity effects, surface tension, internal density differences, and time-periodic forcing.
In the static limit ($a\to 0$), equation~\cref{eqn:Det} reduces to the well-known RT dispersion relation~\citep{chandrasekhar1961hydrodynamic}.

We constrain the displacement of equation~\cref{eqn:Det} to pure sinusoidal harmonic~(H), $\gamma^{(U)} = 0$, or subharmonic~(SH), $\gamma^{(U)} = \rmi \omega/2$, response to obtain the critical acceleration, $a_c$, for neutral stability.
The constraint omits a generalized eigenvalue problem,
\begin{align}\label{eqn:EVD}
  \vb*{A}_{m,\cdot,i}
    \left(
      \gamma_{m,\cdot,i}^{(U)}=0 \quad \text{or} \quad \rmi \omega/2; \, 
      \beta_{m,\cdot,i}
    \right) 
    \hat{\vb*{\zeta}}_{m,\cdot,i} =
      a_c\vb*{B} \hat{\vb*{\zeta}}_{m,\cdot,i},
\end{align}
for the Faraday instability~\citep{kumar1994parametric,kumar1996}.
Equation~\cref{eqn:decomposition_w} also enables the prediction of transient wave dynamics for the azimuthal wavenumber--mode pair $(m,\cdot,i)$ within the unstable regime. 
Previous 2D numerical simulations~\citep{chu2025competing} have shown that the initial transient dynamics can be accurately predicted by 
the modal growth of the most unstable Floquet exponent, $\gamma_{m,\cdot,i}^{(U)}$, combined with the periodic components.
Together, these analyses enable the dissection of the mechanics as the system transitions toward nonlinear interface breakup.

For each azimuthal wavenumber $m$, linear Floquet stability analysis yields a discrete set of radial modes, $\Gamma_{m,\cdot, i}(r)=J_{|m|}(\beta_{m,\cdot ,i} r)/J_{|m|}(\beta_{m,\cdot,i}/ \xi)$, the associated vertical structures, $Z_{m ,n,i}(z)$, and the temporal dependence of each Floquet harmonic,
\begin{align}
  x_{m,n,i}(t) =
        \rme^{\rmi n \omega (t+t_0)+\gamma_{m,\cdot,i} t}.
\end{align}
To enforce incompressibility, we express the perturbation velocity field in poloidal form,
\begin{align}
    \vb*{u}_{m,n,i}(r,z,t)
        = \grad \times \grad \times
        \left(
            x_{m,n,i}(t)\Gamma_{m,n,i}(r) Z_{m,n,i}(z)
            \rme^{\rmi m\theta}\vb*{e}_z
        \right).
\end{align}
Since the radial eigenfunctions are independent of the Floquet harmonic $n$, the velocity components may be rearranged to separate the radial dependence from the $(z,t)$-dependence as
\begin{align}\label{eqn:velocity_cylindrical}
  \mqty[u_r\\
    u_{\theta}\\ w]_{m,\cdot,i} =\rme^{\rmi m\theta}/J_{|m|} (\beta_{m,\cdot,i}/ \xi)
    \mqty[
    \frac{1}{\beta_{m,\cdot,i}}J_{|m|}'(\beta_{m,\cdot,i} r)\\[4pt]
    \frac{\rmi m}{\beta_{m,\cdot,i}^2 r}J_{|m|}(\beta_{m,\cdot,i} r)\\[4pt]
    J_{|m|}(\beta_{m,\cdot,i} r)
    ] \circ 
    \left(\sum_{n=-\infty}^{\infty} x_{m,n,i}(t) \mqty[
    Z'_{m,n,i}\\[4pt]
    Z'_{m,n,i}\\[4pt]
    Z_{m,n, i}
    ](z)\right),
\end{align}
where $\circ$ denotes a Hadamard product.
While the radial eigenfunctions are identical for the two fluids, the axial structure in the linear regime is separated across the unperturbed interface as
\begin{align}
    [Z,\,Z']_{m,n,i}(z) = 
    \begin{cases}
       [Z,\,Z']_{m,n,i}^{(d)}(z), & \text{if } z \leq 0 \\
       [Z,\,Z']_{m,n,i}^{(l)}(z), & \text{if } z > 0.
    \end{cases}
\end{align}
Equation~\cref{eqn:velocity_cylindrical} provides a compact linear spatiotemporal description of the 3-D perturbation field in terms of Floquet harmonics $n$, azimuthal wavenumber $m$, and radial eigenfamilies indexed by $i$.
It therefore enables the prediction of linear-regime dynamics across variations in density contrast, vibration amplitude, and frequency.

\subsection{Pinned contact line at cylinder rim}

In addition to the configuration in which the two fluids are free to move along the cylinder wall, the analysis can be extended to the pinned-contact-line case, for which the interface displacement and therefore the vertical velocity vanish at the cylinder rim, where $r=1/\xi$ and $z=0$:
\begin{align}
    \hat{\zeta}(1/\xi)=\hat{w}(1/\xi,0)=0.
\end{align}
Enforcing this condition requires suitable linear combinations of the radial modes $J_{|m|}(\beta_{m,\cdot, i}r)$~\citep{henderson1994surface,kidambi2013inviscid,shao2021surface,zhang2023pattern}. 
In modal form, evaluating equation~\cref{eqn:total displacement} at $r=1/\xi$ yields the constraint between homo- and inhomogeneous solutions
\begin{align}\label{eqn:pinnedBC_IJ}
    \hat{\zeta}_{m,n,k}^{(\mathrm{H})} = { \sum_{i}\hat{\zeta}^{(\mathrm{P})}_{m,n,i}}. 
\end{align}
Together with equations~\cref{eqn:total displacement,eqn:pinnedBC_IJ}, the total displacement can be written as 
\begin{align}\label{eqn: interface_pinned}
    \hat{\zeta}_{m,n,\cdot} =  \sum_{i}\left(-\lambda_{m,\cdot,i,k}\hat{\zeta}_{m,n, k}^{(\mathrm{H})}+\hat{\zeta}^{(\mathrm{P})}_{m,n,i}\right)\frac{J_{|m|}(\beta_{m,\cdot,i} r)}{J_{|m|}(\beta_{m,\cdot,i}/\xi)}\equiv \sum_i {\hat{\upsilon}_{m,n,i}} \frac{J_{|m|}(\beta_{m,\cdot,i} r)}{J_{|m|}(\beta_{m,\cdot,i}/\xi)}
\end{align}
with the pinned condition $\sum_i {\hat{\upsilon}_{m,n,i}}=0.$
The admissible radial content is then identified by the resulting effective coefficients $\hat{\upsilon}_{m,n ,i}$, which can be written in a compact form as
\begin{equation}\label{eqn:coeffs_Besseli}
\underbrace{{\mqty(\hat{\vb*{\upsilon}}_{m,\cdot,1}^\transpose &
    \hat{\vb*{\upsilon}}_{m,\cdot,2}^\transpose&
    \cdots &
    \hat{\vb*{\upsilon}}_{m,\cdot,N_i}^\transpose
    )}^\transpose}_{\hat{\vb*{\upsilon}}_{m}}=\left(\vb*{I}-\vb*{\Lambda}_k(a)\right)\underbrace{{\mqty(\hat{\vb*{\zeta}}_{m,\cdot,1}^{(\mathrm{P})^\transpose}&
    \hat{\vb*{\zeta}}_{m,\cdot,2}^{(\mathrm{P})^\transpose}&
    \cdots&
    \hat{\vb*{\zeta}}_{m,\cdot,N_i}^{(\mathrm{P})^\transpose}
    )}^\transpose}_{\hat{\vb*{\zeta}}_{m}^{(\mathrm{P})^\transpose}},
\end{equation}
where $\hat{\vb*{\upsilon}}_{m,\cdot,i}$ and $\hat{\vb*{\zeta}}_{m,\cdot,i}^{(\mathrm{P})}$ are the vectors of coefficients over the Floquet harmonics.
The projection coefficient matrix $\vb*{\Lambda}_k(a)$ is obtained from equation~\cref{eqn: lambda}.
In particular, we assume that the total displacement
$\hat{\zeta}_{m,\cdot,\cdot}$ is governed by a single dominant Floquet exponent, $\gamma_{m,\cdot,\cdot}^{(U)}$, such that the exponent associated with the $n$th harmonic component is
$\gamma_{m,n,\cdot}^{(U)}=\gamma_{m,\cdot,\cdot}^{(U)}+ \rmi n\omega$.
Substituting equations~\cref{eqn:pinnedBC_IJ,eqn:coeffs_Besseli} into equation~\cref{eqn:pressure jump_coeffs} leads to a constrained generalized eigenvalue problem
\begin{align}\label{eqn: EVD_pinned_0}
    \left[\vb*{A}^{(\mathrm{P})}_m+\vb*{A}^{(\mathrm{H})}_m\right]\left(\hat{\vb*{\zeta}}_m^{(\mathrm{P})}+\hat{\vb*{\zeta}}_m^{(\mathrm{H})}\right) = a\vb*{B}_m\left(\hat{\vb*{\zeta}}_m^{(\mathrm{P})}+\hat{\vb*{\zeta}}_m^{(\mathrm{H})}\right) \quad \text{subject to} \quad \vb*{W}\hat{\vb*{\zeta}}_m^{(\mathrm{P})}=\vb*{0},
\end{align}
where the block-diagonal matrices
\begin{align}
    \vb*{A}^{[(\mathrm{P}),\,(\mathrm{H})]}_m = \mqty( \vb*{A}^{[(\mathrm{P}),\,(\mathrm{H})]}_{m,\cdot,1} & \vb*{0} & \cdots & \vb*{0} \\
    \vb*{0} &  \vb*{A}^{[(\mathrm{P}),\,(\mathrm{H})]}_{m,\cdot,2} & \ddots &  \vb*{0}\\
    \vb*{0} &   \ddots & \ddots &  \vb*{0}\\
    \vb*{0} &   \cdots & \vb*{0} &  \vb*{A}^{[(\mathrm{P}),\,(\mathrm{H})]}_{m,\cdot,N_i}
    )  \quad \text{and} \quad 
    \vb*{B}_m = \mqty( \vb*{B} & \vb*{0} & \cdots & \vb*{0} \\
    \vb*{0} &  \vb*{B} & \ddots &  \vb*{0}\\
    \vb*{0} &   \ddots & \ddots &  \vb*{0}\\
    \vb*{0} &   \cdots & \vb*{0} &  \vb*{B}
    ),\,
\end{align}
collect the particular and homogeneous contributions and relate the contributions from different radial contents.
The real weight matrix, $\vb*{W}$, imposes the weighted radial-mode condition $\sum_i {\hat{\upsilon}_{m,n,i}}=0$ for each Floquet harmonic.
The matrices $\vb*{A}^{(\mathrm{P})}_{m,\cdot,i}$ and $\vb*{A}^{(\mathrm{H})}_{m,\cdot,i}$ denote, respectively, the particular and homogeneous contributions to the pressure-jump coefficients in equation~\cref{eqn:pressure jump_coeffs} for the $i$th radial mode. 
From equation~\cref{eqn:interface_homo}, the homogeneous operator satisfies
\begin{align}
\vb*{A}^{(\mathrm{H})}_m\hat{\vb*{\zeta}}_m^{(\mathrm{H})} = a\vb*{B}_m\hat{\vb*{\zeta}}_m^{(\mathrm{H})}
\end{align}
Using this together with equation~\cref{eqn:coeffs_Besseli}, equation~\cref{eqn: EVD_pinned_0} reduces to
\begin{align}\label{eqn: EVD_pinned}
    \left[\vb*{A}^{(\mathrm{P})}_m\left(\vb*{I}-\vb*{\Lambda}_k(a)\right)+\vb*{A}^{(\mathrm{H})}_m\right]\hat{\vb*{\zeta}}_m^{(\mathrm{P})} = a\vb*{B}_m\hat{\vb*{\zeta}}_m^{(\mathrm{P})} \quad \text{subject to} \quad \vb*{W}\hat{\vb*{\zeta}}_m^{(\mathrm{P})}=\vb*{0}.
\end{align}
When acting on the particular solution, the homogeneous part of the operator gives
\begin{align*}
    {A}^{(\mathrm{H})}_{m,n,i}
    = 2g_{\text{sgn}} + \frac{\kappa}{\mathrm{Bd}} \beta_{m,n,i}^2 ,
\end{align*}
while the particular part, $\vb*{A}^{(\mathrm{P})}_{m,\cdot,i}$, contains the remaining terms in the coefficient appearing in equation~\cref{eqn:An}.
Projecting the eigenvalue problem onto the constrained subspace, \cref{eqn: EVD_pinned} becomes
\begin{align} \label{eqn: EVD_pinned_subspace}
    \vb*{A}_m'(a) \vb*{v} = a \vb*{B}_m' \vb*{v},
\end{align}
where
\begin{gather}
  \vb*{W}_{\perp} \equiv \text{null}(\vb*{W}), 
  \quad \vb*{A}_m'\equiv\vb*{W}_{\perp}^\transpose\left[\vb*{A}^{(\mathrm{P})}_m\left(\vb*{I}-\vb*{\Lambda}_k(a)\right)+\vb*{A}^{(\mathrm{H})}_m\right]\vb*{W}_{\perp}, 
  \quad \text{and} 
  \quad \vb*{B}_m'\equiv \vb*{W}_{\perp}^\transpose \vb*{B}_m\vb*{W}_{\perp},
\end{gather}
and so the eigenvectors can be recovered as 
\begin{gather}
    \hat{\vb*{\zeta}}_m^{(\mathrm{P})}=\vb*{W}_{\perp}\vb*{v}.
\end{gather}
Equation~\cref{eqn: EVD_pinned_subspace} is solved iteratively under the subharmonic or harmonic response to determine the critical acceleration $a_c$, with the iteration initialized at $a_c=0$.
Given the corresponding eigenvector, the effective coefficient vector
$\hat{\vb*{\upsilon}}_{m}$ is constructed using equation~\cref{eqn:coeffs_Besseli}.
The total interfacial displacement, $\hat{\zeta}_{m,n,\cdot}$, is then recovered from equation~\cref{eqn: interface_pinned}.
Since $\hat{\zeta}_{m,n,\cdot}$ is not associated with a single radial mode number $i$, we characterize the resulting pattern using an effective radial mode number $i^\star$, defined as the number of radial rings in the total interfacial displacement.
The associated velocity field follows as a superposition of retained Floquet harmonics and radial modes,
\begin{align}\label{eqn: velocity field_pinned}
   \left[u_r,\,u_{\theta},\,w \right]_{m}^\transpose= \sum_i\sum_{n=-\infty}^{\infty} \hat{{\upsilon}}_{m,n,i}
    \left[u_r,\,u_{\theta},\,w \right]_{m,n,i}^\transpose.
\end{align}
where the modal velocity field $\vb*{u}_{m,n,i}$ is obtained from equation \cref{eqn:velocity_cylindrical}.
Thus, equation~\cref{eqn: velocity field_pinned} provides a means of predicting the velocity field when the interface is pinned at the cylinder rim.

\section{Results}\label{results}

For demonstration, double-deionized water is chosen as the denser fluid, characterized by density of $\rho^{(d)} = \SI{997}{\kilo\gram\per\meter\cubed}$, kinematic viscosity of $\nu^{(d)} = \SI{1e{-6}}{\meter\squared\per\second}$, and surface tension coefficient of $\sigma^{(d)} = \SI{7.2 e{-2}}{\newton\per\meter}$.
Two lighter-fluid configurations are considered.
The first is air, with density $\rho^{(l)} = \SI{1.20}{\kilo\gram\per\meter\cubed}$ and kinematic viscosity $\nu^{(l)} = \SI{1.51e{-5}}{\meter\squared\per\second}$, giving $\At=0.9976$ and $\eta=\SI{1.81e{-2}}{}$, respectively.
The second is a model fluid specified by $\At=0.5$ and $\eta=0.3$.
The gravitational acceleration is $g = \SI{9.81}{\meter\per\second\squared}$.
\Cref{configurations} summarizes the configurations studied.

\begin{table}[ht!]
    \centering
    \caption{Configurations studied.
    }\label{configurations}
    {\small
    \begin{tabular}{l c c c c c r c}
    Cases & $(H,
    \,R)$~[\unit{\milli\meter}]  &$1/\xi$  & $C $ & $\mathrm{Bd}$ & $(\mathrm{At},\eta)$ & $g_\text{sgn}$ & Section   \\
    \midrule
    Config.~1 &  (22,\,35) &  1.59 & $\SI{9.81e{-5}}{}$   & 65.75 & $\sim(1,0)$ & 1 &  \multirow{2}{*}{\cref{sec:water,sec:results_pinned}}\\
    Config.~2 & (22,\,35)&  1.59 & $\SI{9.81e{-5}}{}$   & 65.75 & $\sim(1,0)$ & $-1$ & \\\midrule
    Config.~3 &  (22,\,35) &  1.59 & $\SI{9.81e{-5}}{}$  & 65.75 & (0.5,0.3) & $-1$ &  \cref{sec:water2} \\\midrule
    Config.~4 &  (38.0,\,27.7) &  0.73 & $\SI{4.32e{-5}}{}$  & 196.2 & $\sim(1,0)$ & $1$ & \cref{sec:results_pinned} \\\midrule
    Config.~5 &  (5.6,\,45) &  8.04 &$\SI{7.62e{-4}}{}$   & 4.26 & $\sim(1,0)$ & $1$ & \cref{sec:results_pinned} \\
    \end{tabular}
    }
\end{table}

We validate the analysis by comparing the predicted growth rates with those of the classical theory for two limiting cases: isolated RT and Faraday instabilities.
Details are provided in \cref{sec:validation}.
In what follows, we consider vibrations without and with static RT instability, denoted by $g_{\text{sgn}}=1$ and $g_{\text{sgn}}=-1$, respectively.
The latter represents an inverted gravitational configuration, in which gravity acts from the denser fluid towards the lighter fluid.
These two cases are referred to as the Faraday-only case and the mixed Faraday--RT case.
In the limit of zero vibration amplitude, the mixed Faraday--RT case reduces to the RT-only configuration.

\subsection{Free-sliding interface}

We first adopt the cylindrical geometry and fluid configuration of the experiments in \citet{shao2021surface} to demonstrate the linear theory predictions.
The cylindrical container has a radius of $R = \SI{0.035}{m}$ and consists of two layers of fluid of equal depth $H = \SI{0.022}{m}$, giving an aspect ratio of $\xi=0.63$.
The dimensionless parameters that quantify the ratios of viscous and capillary forces to gravitational forces are $C = \SI{9.81e{-5}}{}$ and $\mathrm{Bd} = 65.75$, respectively.

\subsubsection{Water--air interface}\label{sec:water}

\begin{figure}[ht!]
  \centering
  \includegraphics{ 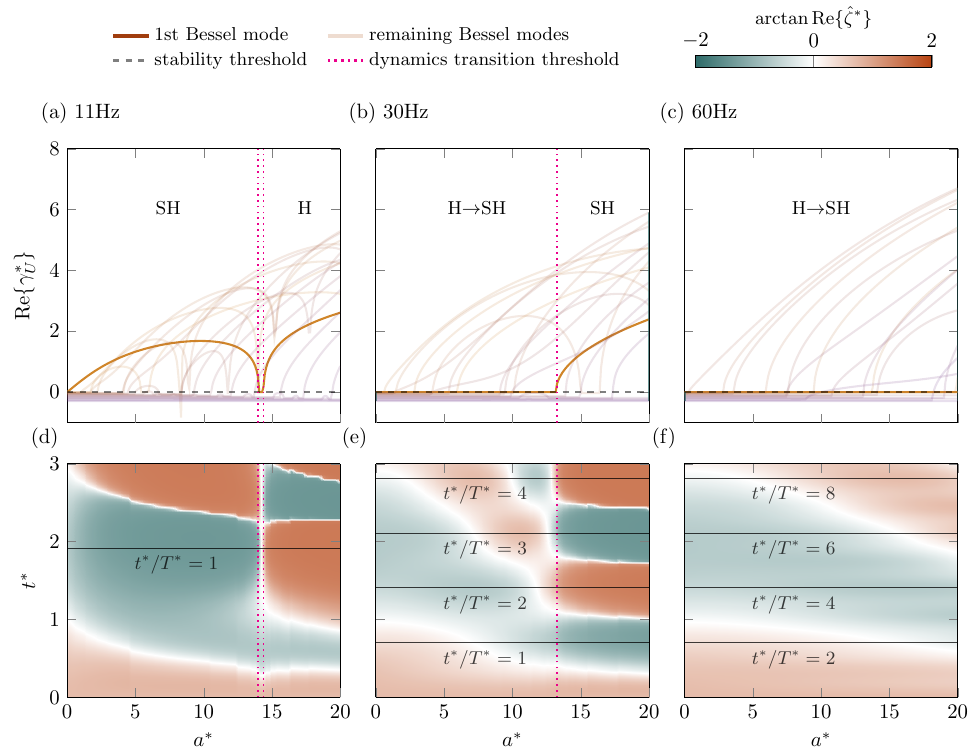}
  \caption{
    Growth rates of the axisymmetric component ($m=0$) for the Faraday-only configuration ($g_{\text{sgn}}=1$) at vibration frequencies as labeled.
    The transitions between subharmonic~(SH, $\mathrm{Im}(\gamma_U^*)= \omega^*/2$) and harmonic~(H, $\mathrm{Im}(\gamma_U^*)=0$) responses are highlighted. 
    The normalized interface displacements for the first Bessel mode $(i=1)$ are also shown.
  }
  \label{fig:Growth_m0_mode1_Faraday} 
\end{figure}

We begin by considering the axisymmetric component ($m=0$) for a water--air interface in the absence of RT instability ($g_{\text{sgn}}=1$).
\Cref{fig:Growth_m0_mode1_Faraday}~(a--c) shows the growth rates for various Bessel modes at three different vibration frequencies. 
The frequency of 11~\unit{\hertz}, shown in panel~\ref{fig:Growth_m0_mode1_Faraday}~(a), corresponds to the natural frequency of the first Bessel mode for the Faraday instability~\citep{shao2021surface}. 
Consistent with this resonance condition, the linear theory predicts a positive growth rate for sufficiently small but finite forcing amplitudes.
As the vibration amplitude increases, higher-order Bessel modes also become unstable and begin to dominate the dynamics.
Concurrently, the first Bessel mode undergoes a transition in its temporal response, shifting from subharmonic~(SH, $\mathrm{Im}(\gamma_U^*)=\omega^*/2$) to harmonic~(H, $\mathrm{Im}(\gamma_U^*)=0$) behavior.
This transition is also reflected in the normalized growth of the first Bessel mode relative to the initial disturbance, as shown in \cref{fig:Growth_m0_mode1_Faraday}~(d--f).
For the harmonic response, the interface displacement retains the same sign after each forcing period, whereas for the subharmonic response, the sign reverses between successive forcing periods.
Here, we plot the arctangent of the normalized displacement to visualize both the exponential amplification and the associated oscillatory phase.
At sufficiently large forcing amplitudes, the amplification is strong enough that the interface quickly enters the nonlinear regime unless the initial disturbance is small.
At higher vibration frequencies, higher-order Bessel modes become unstable first as the vibration amplitude increases.
The first Bessel mode remains stable with harmonic oscillations until a critical bifurcation threshold is reached, beyond which it starts to exhibit subharmonic growth.
\Cref{sec:faraday_freesliding_app} shows the reconstructed linear velocity fields for $\omega=11~\unit{\hertz}$, including the unstable axisymmetric and dipolar responses.

\begin{figure}[ht!]
  \centering
  \includegraphics{ 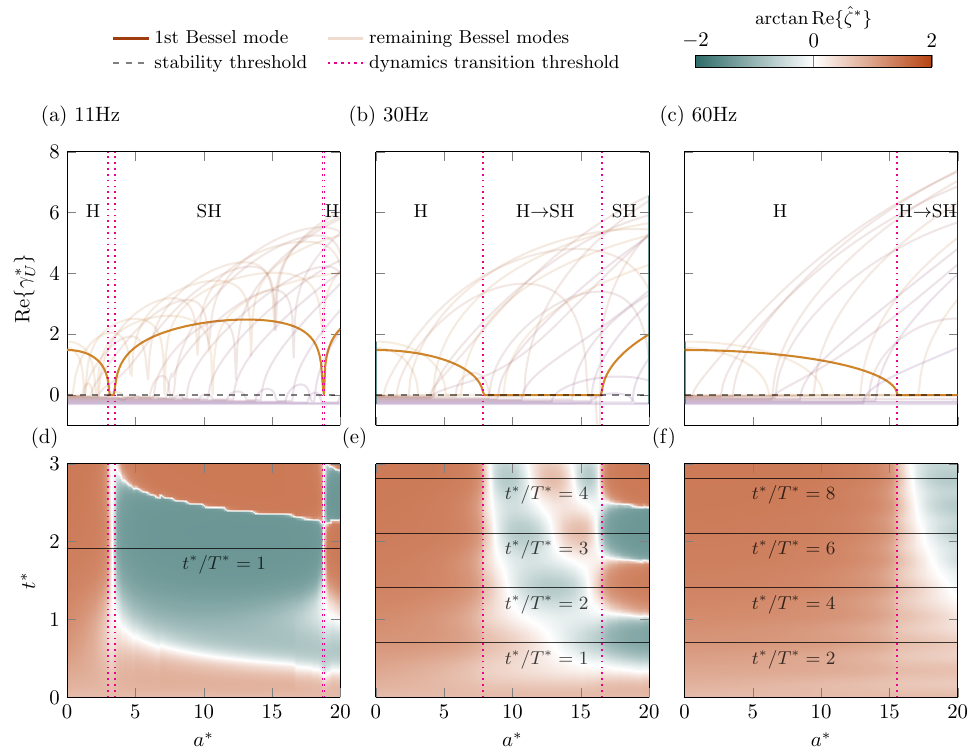}
  \caption{
    Growth rates of the axisymmetric component ($m=0$) for the mixed Faraday--RT configuration ($g_{\text{sgn}}=-1$) at different vibration frequencies: (a) 11~\unit{\hertz}; (b) 30~\unit{\hertz}; and (c) 60~\unit{\hertz}.
    The transitions between subharmonic~(SH, $\mathrm{Im}(\gamma_U^*)=\omega^*/2$) and harmonic~(H, $\mathrm{Im}(\gamma_U^*)=0$) responses are highlighted. 
    The corresponding normalized interface displacements for the first Bessel mode $(i=1)$ are shown in panels (d--f).
   }
  \label{fig:Growth_m0_mode1_FaradayRT} 
\end{figure}

We now consider the mixed Faraday--RT configuration ($g_{\text{sgn}} = -1$) in \cref{fig:Growth_m0_mode1_FaradayRT}.
In the small-amplitude limit ($a\to0$), the first few Bessel modes, which have small radial wavenumbers $\beta$, are unstable, reminiscent of the classical RT instability.
Increasing the vibration amplitude progressively stabilizes these unstable RT modes, while also exciting Faraday-type instabilities.
Each initially unstable RT mode is fully stabilized above a critical acceleration, where its growth rate becomes non-positive.
As the acceleration increases further, the stabilized mode undergoes a Floquet transition in temporal response from H- to SH-type.
This behavior results in a finite acceleration range over which the original RT-type response is stabilized before the subharmonic Faraday response becomes unstable.
At higher vibration frequencies, larger amplitudes are required to stabilize the RT modes, and the corresponding stabilization range broadens.
These observations are consistent with earlier experimental findings~\citep{wolf1969dynamic, wolf1970dynamic} and theoretical analyses of dynamic stabilization~\citep{Troyon1971, lapuerta2001control, piriz2010dynamic, sterman2017rayleigh, pototsky2016faraday} in cylindrical containers.
In the present configuration, however, this range is already occupied by other unstable Faraday modes, leaving the system unstable under oscillatory forcing.
At larger amplitudes, the subharmonic Faraday response further transitions to a harmonic Faraday regime.
Increasing the vibration amplitude, therefore, produces a sequence of modal responses: a low-amplitude RT-dominated harmonic regime, a dynamically stabilized interval, a subharmonic Faraday-dominated regime, and, under stronger forcing, a harmonic Faraday-dominated regime.
This sequence is most clearly observed in \cref{fig:Growth_m0_mode1_FaradayRT}~(a) at lower vibration frequency, whereas higher frequencies require a wider amplitude range to reach the same transitions.
Although both the initial RT response and the subsequent harmonic Faraday response have purely real Floquet exponents, their temporal dynamics are distinct.
\Cref{fig:Growth_m0_mode1_FaradayRT}~(d--f) shows the periodically oscillating components, $\hat{\zeta}_{m,n, i}$, which induce finite phase variations in the harmonic Faraday response.
In the RT response, the components instead modulate the steady exponential growth without changing the dominant dynamics.

\begin{figure}[ht!]
  \centering
  \includegraphics{ 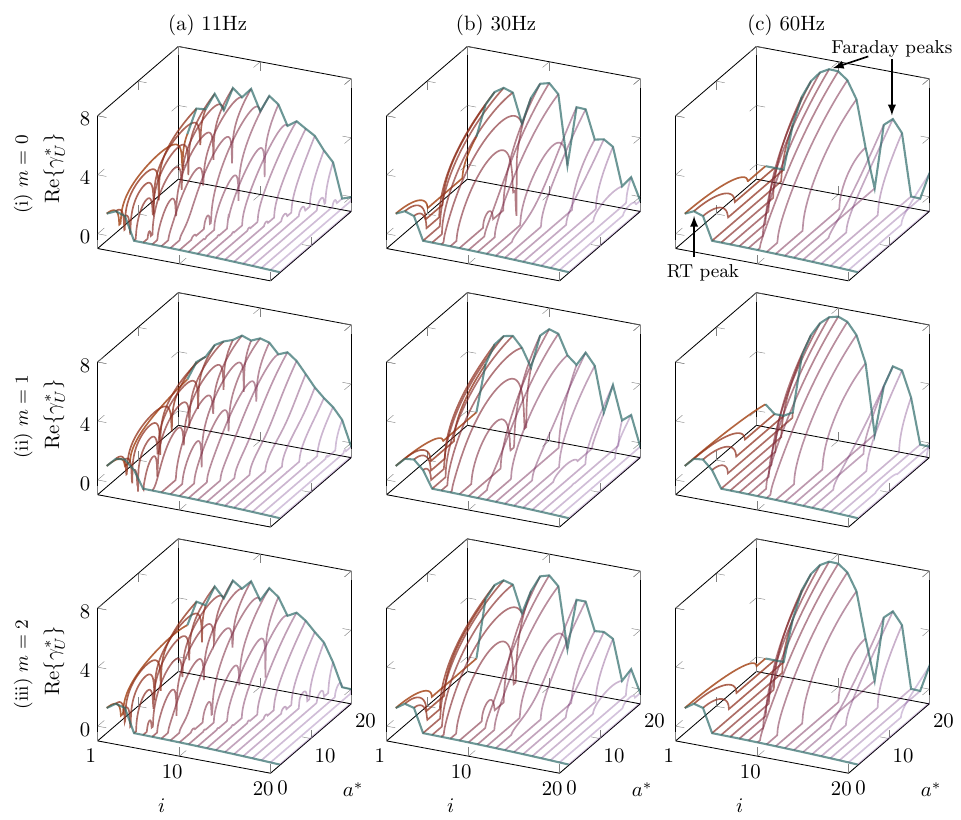}
  \caption{
    Growth rates of the axisymmetric ($m=0$; i), dipolar ($m=1$; ii) and quadrupolar ($m=2$; iii) modes for the mixed Faraday--RT configuration ($g_{\text{sgn}}=-1$) at vibration frequencies as labeled.
  }
  \label{fig:Growthrate_FaradayRT} 
\end{figure}

To examine the instability mechanisms in the mixed Faraday--RT configuration, we compare the growth rates across azimuthal wavenumbers, vibration frequencies, and amplitudes in \cref{fig:Growthrate_FaradayRT}; the corresponding Faraday-only results are shown in \cref{fig:Growthrate_Faraday}.
The general trends are similar for the axisymmetric ($m=0$), dipolar ($m=1$), and quadrupolar ($m=2$) modes: vibration can stabilize initially unstable RT modes at low radial mode indices while simultaneously exciting Faraday modes at higher mode indices.
External vibration thus excites not only radial modes but also higher-order azimuthal modes through the same mechanisms.
Consequently, for initial disturbances containing appreciable energy in multiple azimuthal components, strong nonlinear mode interactions are anticipated, especially when several modes have comparable growth rates.
The discreteness of the wavenumber-like parameter $\beta_{m,\cdot, i}$ in cylindrical geometry also leads to mode-dependent growth rates among adjacent Bessel modes, indicating that neighboring modes respond differently to the imposed vibration.
At higher vibration amplitudes, these modal differences become less pronounced as the Faraday instability dominates, yielding more uniform responses across adjacent modes.
Higher-order Faraday peaks emerge beyond the fundamental one.
These mode-dependent effects are most pronounced at lower vibration frequencies, where Faraday and RT mechanisms coexist, and induce complex interfacial dynamics when several unstable modes are excited simultaneously.

\begin{figure}[ht!]
  \centering
  \includegraphics{ 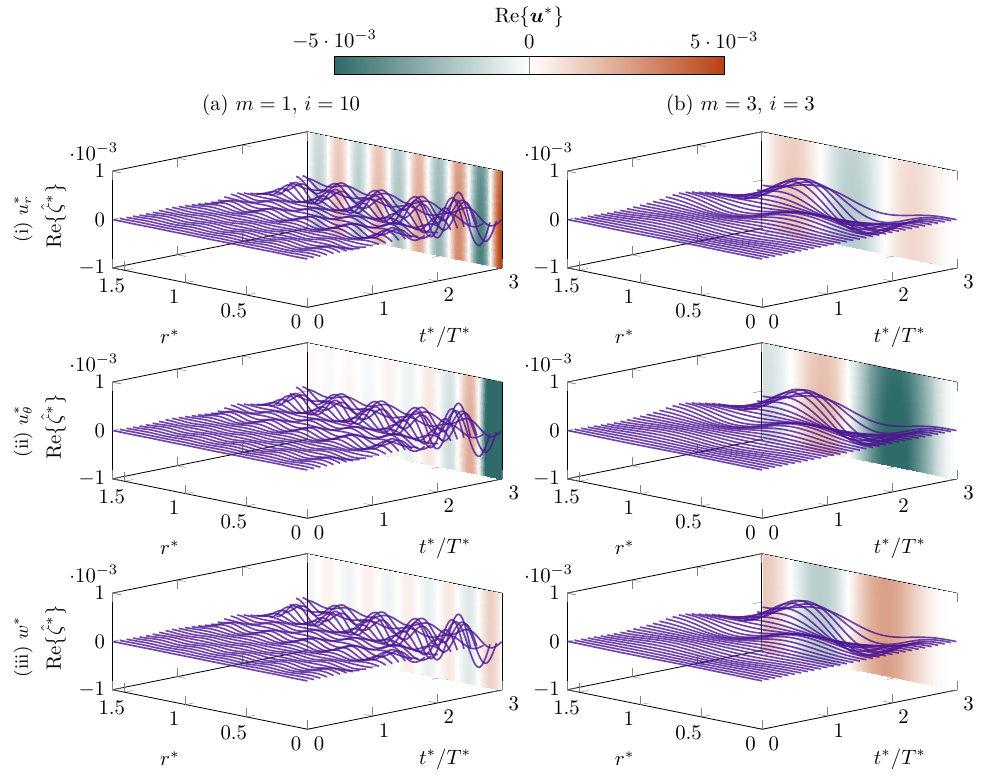}
  \caption{
      Predictions of the interfacial displacement and velocity fields at $t^*/T^*=3$ for the mixed Faraday--RT configuration ($g_\text{sgn}=-1$) at vibration amplitude $a^*=6.3$ and frequency $\omega=30~\unit{\hertz}$: (a) $m=1, i=10, \gamma_U^*=1.783$; (b) $m=3$, $i=3$, $\gamma_U^*=1.663+\rmi\omega^*/2$.
    }
    \label{fig:Interface_velocity_Faraday}
\end{figure}

\Cref{fig:Interface_velocity_Faraday} shows the predicted spatiotemporal interface dynamics and flow field for the mixed Faraday--RT configuration.
As a representative case, we consider a forcing amplitude $a^*=6.3$ and frequency $\omega=\SI{30}{\hertz}$.
Under these conditions, the dipolar mode exhibits the unstable harmonic Faraday response shown in \cref{fig:Growthrate_FaradayRT}, whereas the omitted hexapolar ($m=3$) mode exhibits an unstable subharmonic response.
Although these two modes have comparable growth rates, their resulting interfacial dynamics differ markedly.
The dipolar mode exhibits substantially larger spatiotemporal oscillations than the hexapolar mode, consistent with its harmonic response, for which the interface returns to the same phase after each forcing period.
By contrast, the subharmonic hexapolar response evolves over two forcing periods, leading to a weaker apparent oscillation over the single-period visualization shown here.
The two modes also differ in the spatial structure of the induced velocity field: the dipolar mode exhibits a stronger radial velocity component, whereas the hexapolar mode shows a relatively larger axial velocity amplitude.
In both modes, the largest velocity amplitude occurs in the azimuthal component, indicating strong azimuthal motion in these non-axisymmetric responses.

\begin{figure}[ht!]
  \centering
  \includegraphics{ 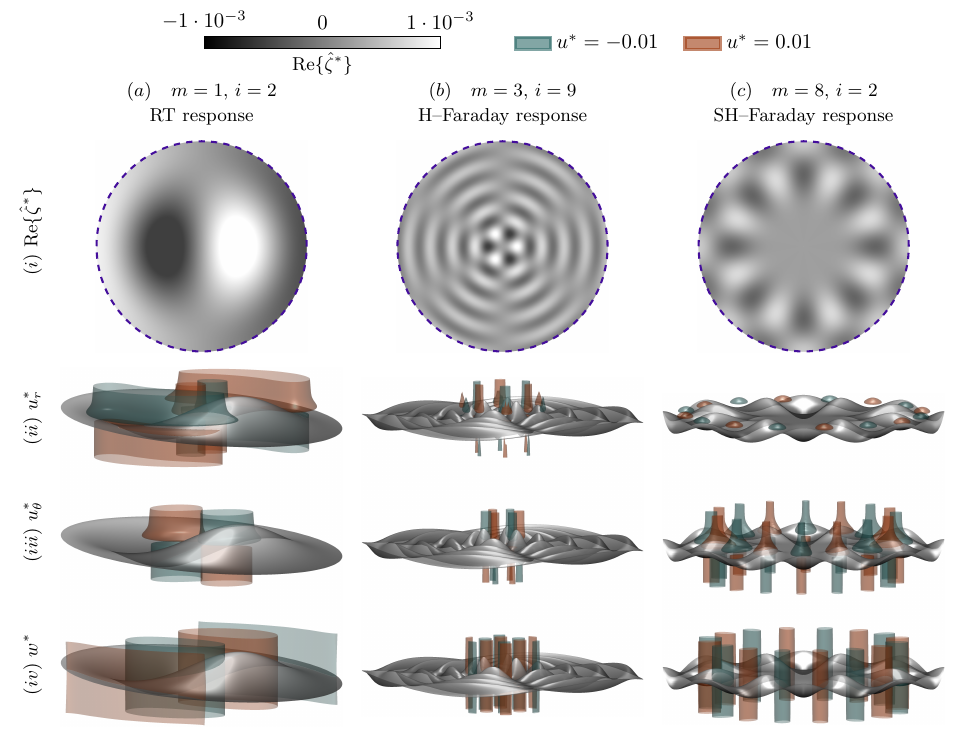}
  \caption{
    Predictions of the instantaneous velocity fields and corresponding interfacial deformations for the mixed Faraday--RT configuration ($g_\text{sgn}=-1$) at vibration amplitude $a^*=6.3$ and frequency $\omega=30~\unit{\hertz}$: (a) $m=1, i=2$; (b) $m=3,\, i=9$; and (c) $m=8,\, i=2$.
    Different instability responses are triggered for these three cases.
    The fields are shown over the vertical range $z^*\in[-2,2]\times10^{-3}$.
  }
  \label{fig:Interface_velocity_Faraday_RT}
\end{figure}

For the same configuration in \cref{fig:Interface_velocity_Faraday}, \cref{fig:Interface_velocity_Faraday_RT} shows the coupling between the instantaneous velocity fields and the corresponding interface deformation.
For demonstration, results are shown for three azimuthal wavenumbers, representing RT-dominated, harmonic, and subharmonic Faraday responses, respectively.
Over the vertical range considered, only the radial and azimuthal velocity components exhibit sign changes across the interface, indicating that the two fluids move coherently in the axial direction while developing lateral interfacial shear.
Consistent with \cref{fig:Growthrate_FaradayRT}, where the RT response is confined to low radial mode indices, the RT-dominated mode has a larger-scale spatial structure than the Faraday modes.
As the azimuthal wavenumber increases from $m=1$ to $m=8$, the second radial mode transitions from an RT-dominated response to a subharmonic Faraday response, with the largest interface deformation and velocity amplitudes localized near the cylinder wall.
For the harmonic Faraday mode at $m=3$ and $i=9$, the largest deformation instead occurs near the cylinder axis, leading to more compact induced velocity fields.

\subsubsection{Water--model-fluid interface}\label{sec:water2}

\begin{figure}[ht!]
  \centering
  \includegraphics{ 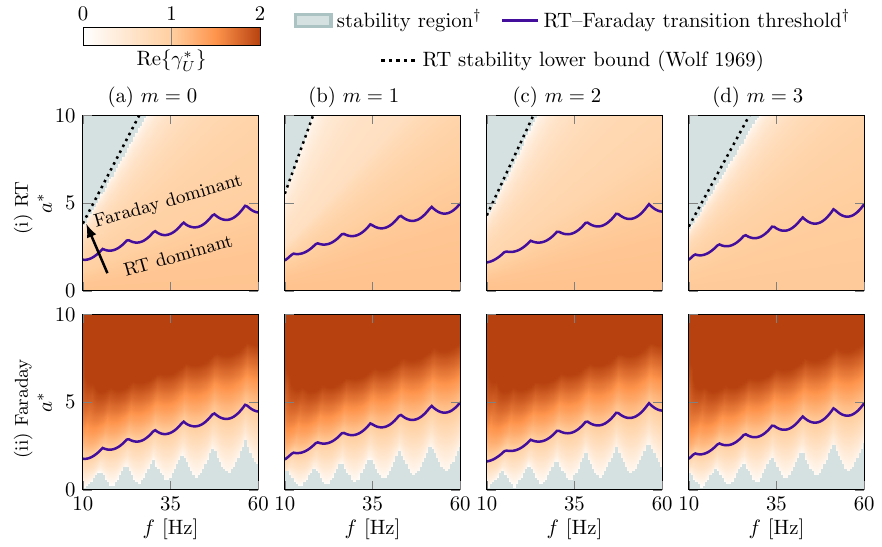}
  \caption{
    Dominant growth rates associated with (i) RT- and (ii) Faraday-type instabilities across the vibration frequency--amplitude phase space for different azimuthal wavenumbers: (a) $m=0$; (b) $m=1$; (c) $m=2$; and (d) $m=3$. 
    The present analysis$^\dagger$ identifies the RT- and Faraday-stability regions, together with the margins of the Faraday--RT transition.
    For comparison, the lower bounds for RT dynamic stabilization predicted by \citet{wolf1969dynamic} are also shown.
  }
  \label{fig:phase space_case2}
\end{figure}

Next, we examine the competition between RT and Faraday instabilities for a water--model-fluid interface, characterized by $\At=0.5$ and $\eta=0.3$.
This moderate Atwood number and viscosity ratio provide a clearer regime for studying instability competition at a liquid--liquid interface.
\Cref{fig:phase space_case2} shows the dominant growth rates associated with RT and Faraday instabilities across the vibration frequency--amplitude space for different azimuthal wavenumbers.
We discretize the vibration parameter space with resolutions $\Delta f= \SI{0.625}{\hertz}$ and $\Delta a^*= 0.125$.
We distinguish the two instabilities by their origins in the growth-rate--wavenumber spectra shown in \cref{fig:validation_case2}.

Overall, increasing the vibration amplitude $a^*$ suppresses the RT growth rate and dynamically stabilizes the RT instability beyond a critical threshold.
The resulting stability boundary agrees favorably with the threshold relation of \citet{wolf1969dynamic}.
The inclusion of viscous and surface-tension effects in the present analysis slightly expands the predicted stability region relative to this simplified threshold estimate.
Complete stabilization relies on confinement~\citep{chu2025competing}, which we show in \cref{sec:confinement_app}.
For the present fluid configuration and range of vibration parameters, the most unstable Faraday modes are all subharmonic.
The stability boundaries of the Faraday response in \cref{fig:phase space_case2}~(ii) clearly delineate the Faraday tongues.
Complete stabilization of the interfacial instability can occur only when the RT-stable and Faraday-stable regions overlap.
Such an overlap is not observed here; it typically requires much higher viscosities~\citep{wolf1970dynamic}.

As the vibration amplitude increases further, the Faraday growth rates generally increase and eventually exceed those of the RT modes.
The transition boundary from RT- to Faraday-dominated behavior does not vary monotonically with vibration frequency, but instead follows the structure of the Faraday tongues.
These results reveal a four-stage transition in the dominant response.
At low vibration amplitudes, RT waves dominate the interfacial dynamics, and the Faraday instability is not yet excited.
Once the corresponding Faraday threshold is exceeded, Faraday waves acquire a positive growth rate, although the leading RT mode remains dynamically dominant.
With increasing vibration amplitude or decreasing frequency, the leading RT growth rate decreases, while the leading Faraday growth rate increases, leading to a transition to Faraday-dominated dynamics.
For vibration amplitudes exceeding the RT dynamical-stabilization threshold, the initially RT-unstable modes are stabilized in the confined domain.

\begin{figure}[ht!]
  \centering
  \includegraphics{ 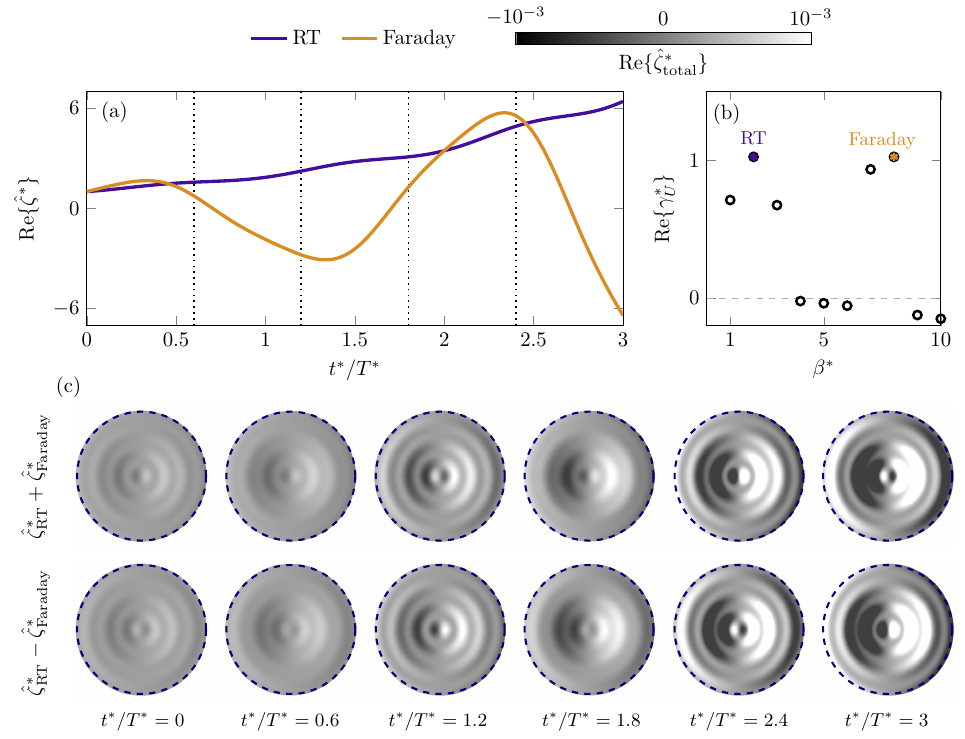}
  \caption{
    Evolution of the dipolar modes ($m=1$) for $f=\SI{30}{\hertz}$ and $a^*=3.71$: (a) normalized Faraday and RT modes; (b) corresponding Floquet growth rates; and (c) combined interfacial patterns from in-phase and out-of-phase superpositions of the RT and Faraday modes.
  }
  \label{fig:interfacial dynamics_FaradayRT}
\end{figure}

As an example, \cref{fig:interfacial dynamics_FaradayRT} shows the interfacial dynamics of the dipolar modes when the RT and Faraday modes have identical growth rates.
The RT mode grows steadily, superposed with a periodic modulation.
In contrast, the subharmonic Faraday mode reverses sign during the first forcing period and subsequently rebounds to a positive peak, intersecting the RT displacement at $t^*/T^*=2$.
Depending on their relative initial phases, $t_0^*$, the RT and Faraday modal amplitudes can follow different trajectories within a forcing period.
Superposing the two modes can produce distinct interfacial patterns.
Here, we show the temporal evolution of two representative patterns over three forcing periods, formed by the sum and difference of the RT and Faraday modes and corresponding to in-phase and out-of-phase superpositions, respectively.
Constructive and destructive interference between the two responses yields distinct instantaneous interfacial deformations, suggesting that the interface pattern can depend strongly on the modal composition of the initial disturbance.

\subsection{Pinned contact line}\label{sec:results_pinned}

We next turn to the pinned-contact-line case.
While several experimental studies have focused on the Faraday-only configuration, mixed Faraday--RT regimes remain less well understood.
We first validate the present linear Floquet stability analysis for a water--air interface against existing Faraday-only experimental measurements, and then extend the analysis to investigate the competing instability mechanisms in mixed Faraday--RT regimes.

\subsubsection{Faraday-only configuration}

\begin{figure}[ht!]
  \centering
  \includegraphics{ 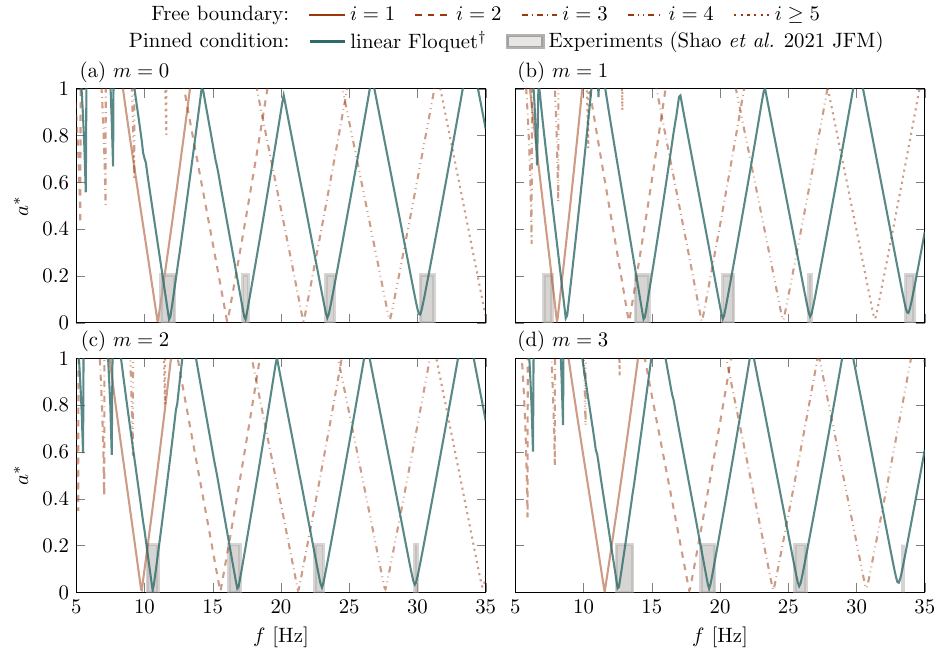}
  \caption{
    Comparison between the critical frequency predicted by the present linear Floquet stability theory, $f_{\mathrm{LFST}}^\dagger$, and the experimental measurements of \citet{shao2021surface} for the Faraday-only case ($g_\text{sgn}=1$): (a) $m=0$; (b) $m=1$; (c) $m=2$; and (d) $m=3$.
    Corresponding critical accelerations obtained without the pinned-condition constraint are shown for comparison.
  }
  \label{fig:Resonance tongues_Faraday}
\end{figure}

\Cref{fig:Resonance tongues_Faraday} compares the critical frequency predicted by the present linear Floquet stability theory, $f_{\mathrm{LFST}}^\dagger$, and the experimental measurements of \citet{shao2021surface} for the Faraday-only configuration ($g_\text{sgn}=1$).
The corresponding cylindrical geometry and fluid properties are summarized in \cref{configurations}.
In the experiments, the critical interfacial patterns are not observed at zero acceleration; instead, a finite vibration acceleration of approximately $\SI{2}{\meter\per\second\squared}$ is required to trigger the instability patterns.
Across the considered azimuthal wavenumbers, the measured onset frequencies lie close to the predicted critical frequencies, within a vibration-amplitude threshold of $0.2$.
This agreement validates the present theory, which accounts for viscous effects, the two-fluid interface, and the Floquet-type temporal response.
Enforcing the interface profile to be pinned at the cylinder rim increases the onset frequency required to excite the instability patterns compared with the corresponding free-sliding configuration.
This increase is expected because the critical frequency increases with radial mode index, and satisfying the pinned condition requires a superposition involving higher radial modes.
Within the frequency range $\SIrange{5}{10}{\hertz}$, additional local minima are observed in the critical-acceleration curves.
These features can be interpreted as combination resonance tongues \citep{kidambi2013inviscid,zhang2023pattern}, indicating resonant coupling among different radial modes under parametric forcing.

\begin{table}[ht!]
    \centering
    \caption{Complementary comparison of the predicted critical vibration frequency, $f_{\mathrm{LFST}}^\dagger$, with previous experimental measurements, $f_{\mathrm{exp}}$, for the Faraday-only configuration ($g_\text{sgn}=1$).
    }\label{exp_comparisons}
    {\small
    \begin{tabular}{r @{\hspace{1em}} c c c c c c l}
    Experiment & $h$~[\unit{\milli\meter}] & $R$~[\unit{\milli\meter}]  &$(m,i^{\star})$  & $f_{\mathrm{exp}}~[\unit{\hertz}] $ & $f^\dagger_{\mathrm{LFST}}~[\unit{\hertz}]$ & $\Delta f$&   \\
    \midrule
    \multirow{6}{*}{\citet{henderson1994surface}} & \multirow{6}{*}{38.0} & \multirow{6}{*}{27.7} &   (0,1)  & 13.52 & 13.68 & $+1.18\%$\\
      &  &  &    (1,1)  & \phantom{1}9.30 & \phantom{1}9.46 & $+1.72\%$ \\
      &  &  &    (1,2)  & 17.20 & 17.30 & $+0.58\%$\\
      &  &  &    (2,1)  & 12.71 & 12.64 & $-0.55\%$\\
      &  &  &    (3,1)  & 15.68 & 15.60 & $-0.51\%$\\
      &  &  &    (4,1)  & 18.58 & 18.52 & $-0.32\%$\\ \midrule
    \multirow{6}{*}{\citet{shao2021surface}} &  \multirow{6}{*}{22} &  \multirow{6}{*}{35} &  (0,5) &37.9   & 37.70 & $-0.53\%$ \\
    &   &   &  (1,6) &42.0   & 41.68 & $-0.76\%$ \\
    &   &   &  (5,1) &18.1   & 17.48 & $-3.43\%$ \\
    &   &   &  (5,2) &25.1   & 24.43 & $-2.67\%$ \\
    &   &   &  (6,1) &19.9   & 19.68 & $-1.11\%$ \\
    &   &   &  (6,2) &27.6   & 27.08 & $-1.88\%$ \\ \midrule
   \multirow{3}{*}{ \citet{zhang2023pattern}} & \multirow{3}{*}{5.6} & \multirow{3}{*}{45}   & (2,3) & 15.5 & 16.55 & $+6.77\%$\\
   &  &    & (3,3) & 18.5 & 19.25 & $+4.05\%$\\
   &  &   & (4,2) & 15.2 & 15.82 & $+4.08\%$
    \end{tabular}
    }
\end{table}

\Cref{exp_comparisons} presents complementary comparisons to those in \cref{fig:Resonance tongues_Faraday}, covering additional azimuthal wavenumbers reported by \citet{shao2021surface} and different cylindrical geometries by \citet{henderson1994surface} and \citet{zhang2023pattern}.
The predicted critical frequencies agree well with the experimental measurements, with deviations below $2\%$ in most comparisons.
Although larger discrepancies are observed for the high-aspect-ratio configuration of \citet{zhang2023pattern}, the predictions still capture the measured onset frequencies with reasonable accuracy.
Overall, these comparisons support the validity of the present analysis for the Faraday-only configuration across the range of cases considered.
These experiments relied on high-speed imaging of the top-view interfacial deformation and therefore did not provide direct access to the corresponding instantaneous velocity fields.
The present analysis complements such measurements by predicting both the critical vibration conditions and the associated velocity field within the linear regime, enabling a more complete spatiotemporal description of the interfacial dynamics.

\begin{figure}[ht!]
  \centering
  \includegraphics{ 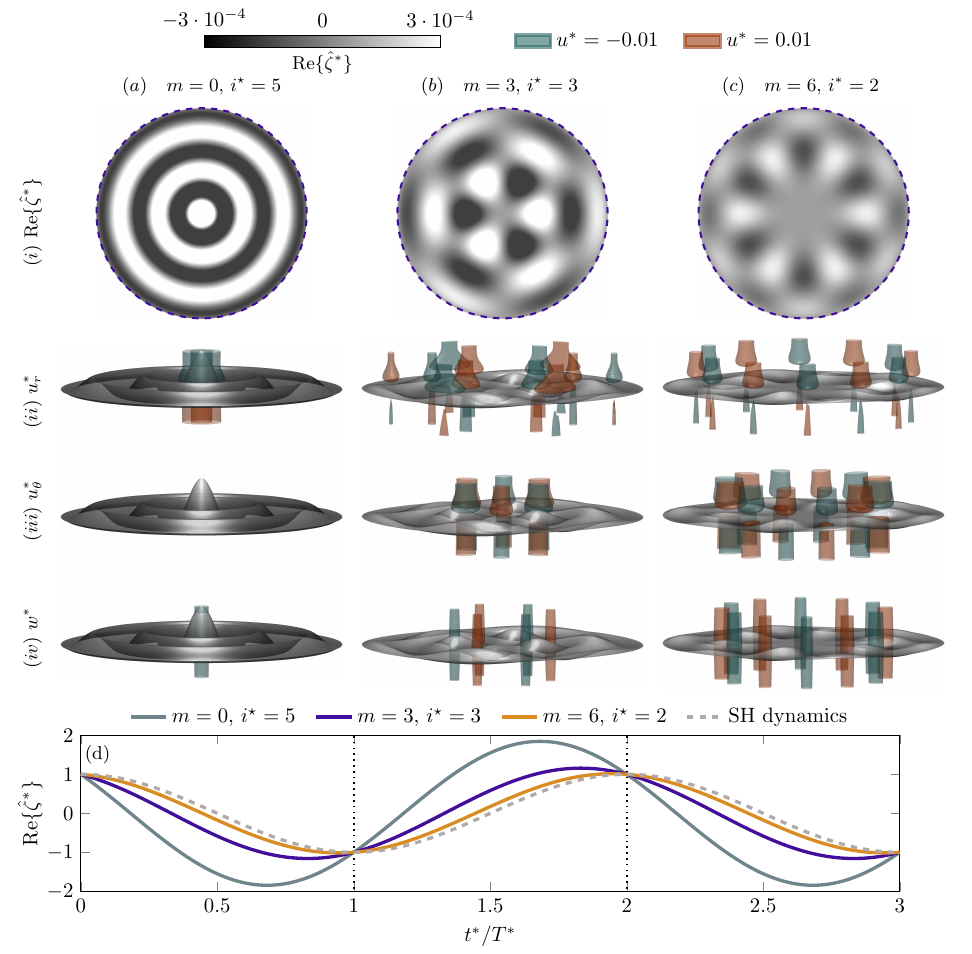}
  \caption{
    Predictions of interfacial deformations and the corresponding velocity fields for the Faraday-only configuration ($g_\text{sgn}=1$) with a pinned contact line at the cylinder rim.
    Results are shown at the respective critical frequencies for (a) $m=0, i^\star=5$; (b) $m=3,\, i^\star=3$; and (c) $m=6,\, i^\star=2$, with the corresponding normalized temporal evolutions shown in (d).
    The velocity fields are shown over the vertical range $z^*\in[-2,2]\times10^{-3}$.
  }
  \label{fig:Interface_velocity_Faraday_pinned}
\end{figure}

\Cref{fig:Interface_velocity_Faraday_pinned} shows three representative examples of interfacial deformations and their corresponding predicted velocity fields.
The selected interfacial patterns are consistent with those observed experimentally~\citep{shao2021surface,zhang2023pattern}.
The responses are evaluated at their respective critical frequencies, where the modes are neutrally stable at the onset of the Faraday instability.
The instantaneous velocity fields nevertheless reveal the spatial structure of the neutral modes, including lateral interfacial motion and the associated shear near the interface.
The axial velocity fields are similar to those obtained for the free-sliding interface in \cref{fig:Interface_velocity_Faraday_RT}, with the largest velocity amplitudes occurring near the regions of maximum interfacial displacement.
The accompanying Floquet harmonics modify the dominant subharmonic response, leading to mode-dependent temporal evolution of the interfacial patterns.
In short, the pinned-contact-line condition constrains the near-rim interfacial motion, increases the critical onset frequencies, and redistributes the lateral flow along the interface.

\subsubsection{Mixed Faraday--RT configuration}

\begin{figure}[ht!]
  \centering
  \includegraphics{ 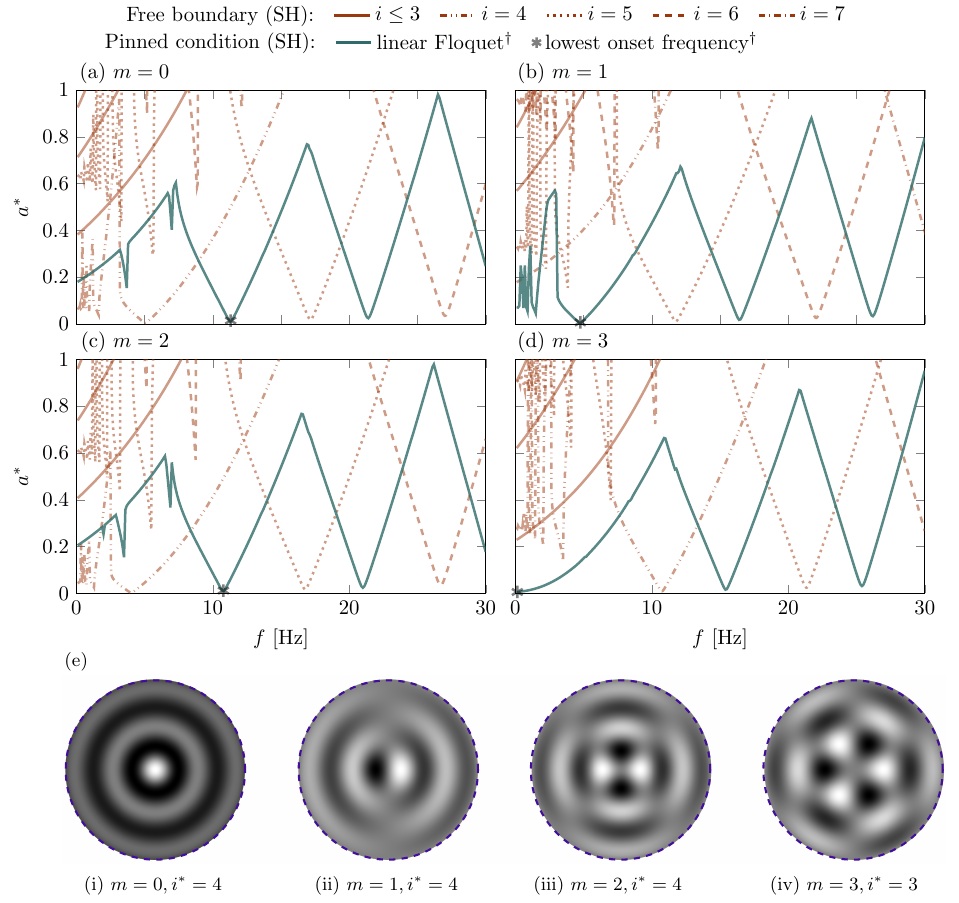}
  \caption{
    Critical frequency for the mixed Faraday--RT configuration ($g_\text{sgn}=-1$): (a) $m=0$; (b) $m=1$; (c) $m=2$; and (d) $m=3$.
    Panel (e) shows the interfacial patterns associated with the lowest onset frequency.
    Critical acceleration for the free-sliding case is included for comparison.
    }
    \label{fig:Resonance tongues_FaradayRT}
\end{figure}

We next consider a more complex configuration in which the interface remains pinned at the cylinder rim but is subject to static RT instability ($g_{\text{sgn}}=-1$).
In the absence of imposed vibration, such a nominally RT-unstable configuration may be sustained in a non-Newtonian fluid--air system, where elastic stresses can suppress the interfacial growth~\citep{zheng2019rayleigh,bret2026linear}.
Alternatively, for Newtonian fluids, a comparable neutrally stabilized state can instead be realized dynamically through vertical vibration.
This configuration differs fundamentally from the Faraday-only case in \cref{fig:Resonance tongues_Faraday}, as the pinned condition requires a superposition of unstable RT and stable Faraday modes to enforce zero displacement at the cylinder rim.
Since the present analysis focuses on the critical acceleration for neutral stability, we assume that the admissible radial structure can be represented by a single Floquet exponent with zero real part, together with its associated Floquet harmonics.

Accordingly, \cref{fig:Resonance tongues_FaradayRT} shows the predicted critical acceleration as a function of vibration frequency for a water--air interface in the mixed Faraday--RT configuration.
In the absence of the pinned-contact-line constraint, the first few radial modes for the considered azimuthal wavenumbers ($i\leq3$ for $m=0,2,3$ and $i\leq 4$ for $m=1$) are unstable without vibration owing to the RT mechanism.
In the quasi-static limit $f\to0$, the stabilization thresholds of these RT modes satisfy $a^*<1$, indicating that stability can be reached before the effective gravity is fully removed, owing to the finite-wavelength capillary restoring force and additional viscous damping.
The RT stabilization thresholds increase with vibration frequency.
The higher modes have radial wavelengths larger than the RT cutoff and exhibit Faraday-type behavior: they are stable at low accelerations and become unstable only above their respective critical values.
Consequently, near-zero critical accelerations can arise only for combined modes, in which unstable RT and stable Faraday components are superposed to satisfy the zero-displacement condition at the cylinder rim.
This modal superposition produces a much richer radial structure at onset, with more than three radial rings appearing already at the lowest onset frequency; see \cref{fig:Resonance tongues_FaradayRT}~(e).
In contrast, when unstable RT modes are absent, the onset mode consists of only a single radial ring~\citep{shao2021surface}.
While the other azimuthal wavenumbers retain finite lowest onset frequencies, the $m=3$ branch approaches the quasi-static limit $f\to0$, indicating a neutral balance between the RT-unstable components and the stable higher-radial components.
Together, the analysis predicts that the pinned-contact-line constraint does more than shift the onset frequencies.
Rather, it qualitatively alters the onset mechanism by enforcing interactions between RT-unstable and Faraday-stable modes.

\section{Discussion and conclusion}
\label{discussion}

\subsection{Limitations and outlook}

The present Floquet analysis is linear and therefore describes the onset and early growth of the interfacial instabilities rather than their nonlinear evolution.
In the linear problem, the rotational symmetry of the base state lets disturbances with different azimuthal wavenumbers evolve independently.
Beyond onset, the convective terms in the Navier--Stokes equations couple these modes, transferring momentum and kinetic energy among azimuthal wavenumbers, radial modes, and Floquet harmonics through sum- and difference-type interactions.
Because the radial projection of these nonlinear products onto the Bessel basis need not vanish, several radial modes may be excited simultaneously, producing interfacial responses that a single linear Floquet mode cannot represent.
Interactions among linearly stable modes can also provide finite-amplitude forcing to linearly unstable ones, modifying the early post-onset dynamics.
Such modal interactions and pattern evolution have been observed experimentally for the Faraday-only configuration~\citep{zhang2023pattern}, but we know of no corresponding theoretical description.
Weakly nonlinear analysis offers a route to representing these effects, as demonstrated for Faraday-only configurations in two-dimensional domains~\citep{zhang1996square,chen1997pattern,chen1999pattern} and for axisymmetric dynamics~\citep{bongarzone2022subharmonic}.
Extending such approaches to include azimuthal mode interactions in the presence of RT instability is a natural future direction.

For pinned contact lines, the admissible interface is constructed from a superposition of radial modes that share a Floquet exponent.
This representation is self-consistent for the onset predictions that we focus on: at the critical acceleration, the collective mode is neutrally stable, so a common exponent, harmonic ($\gamma_U=0$) or subharmonic ($\gamma_U=\rmi\omega/2$), is determined by the neutral-stability condition, not assumed.
Approximation error arises only away from the onset, where the coupled radial modes can acquire distinct growth rates and temporal symmetries, particularly when associated with different instability mechanisms, so the single-exponent structure no longer fully describes the instability's evolution.
The pinned condition is idealized: we impose a fixed contact line and neglect contact-angle hysteresis and dynamic contact-angle effects.
Experimentally realizing the mixed Faraday--RT regime is challenging, as vibration, an adverse density contrast, and a pinned contact line must be imposed simultaneously, and the contact line may still depin intermittently or have phase-dependent movement.
Thus, representing the nonlinear saturation and mode interactions beyond onset requires direct simulation to determine how the coupled Faraday--RT modes nonlinearly evolve under the pinned constraint.

\subsection{Conclusions}
\label{conclusion}

This study presents a theoretical investigation of the Rayleigh--Taylor (RT) and Faraday instabilities in vertically vibrated cylindrical containers, with particular focus on their coexistence and on additional linear coupling mediated by a pinned-contact-line constraint.
Each mechanism is fundamental to hydrodynamic instabilities and has been extensively studied in isolation; their coupled behavior in a confined cylindrical geometry remains less well understood, especially when the interface is constrained at the container rim.
Using linear Floquet stability analysis, we characterize the onset and dynamics of the instabilities across the vibration frequency--amplitude parameter space, resolved by azimuthal wavenumber, Floquet harmonic, and radial (Bessel) mode index.
The analysis shows that RT dynamics are dominated by long-wavelength modes that grow steadily, whereas Faraday responses preferentially excite shorter-wavelength modes with harmonic or subharmonic Floquet time dependence.
The formulation recovers the classical dispersion relations in the limiting cases of RT-only and Faraday-only instability.
It is further validated against experimental measurements for both the dynamical stabilization of RT waves and the Faraday instability under the pinned-contact-line constraint.

In the mixed Faraday--RT regime, where both external vibration and internal density contrast are present, we show that the pressure-gradient-driven RT instability and the parametrically excited Faraday instability can coexist and compete under the free-sliding boundary condition.
For a prescribed vibration frequency, we show that increasing the vibration amplitude can produce a sequence of modal responses among the RT-unstable radial modes: a low-amplitude harmonic, RT-dominated response; a transition beyond a critical amplitude to a subharmonic Faraday regime; and, under stronger forcing, a further transition to a harmonic Faraday regime.
For each azimuthal wavenumber, we further examine the instability mechanism associated with the most unstable radial modes.
We identify a four-stage transition in the dominant response: an RT-dominated regime at low forcing; the onset of positive-growth Faraday waves while RT remains dominant; a crossover to Faraday-dominated dynamics as RT growth is suppressed; and, finally, dynamic stabilization of the initially RT-unstable modes in the confined domain.

When the pinned-contact-line constraint is imposed, it enforces zero displacement at the cylinder rim and thereby couples radial modes that would otherwise evolve independently.
In the Faraday-only regime, this modal superposition shifts the onset to higher critical forcing frequencies relative to the free-sliding case.
In the mixed Faraday--RT regime, the pinned condition has a more profound effect.
The admissible onset modes require a superposition of unstable RT components and stable Faraday components to satisfy the boundary constraint.
This mechanism suggests a vibration-induced analog of the stabilization that elastic stresses provide in RT-unstable non-Newtonian fluid--air systems.
Within linear theory, therefore, the pinned contact line does more than modify the instability threshold; it qualitatively changes the onset mechanism by mediating linear coupling between RT- and Faraday-type modes.
Direct simulation can now test this parsimoniously.

Lastly, the analysis extends beyond the interfacial patterns to the associated linear flow fields, which are challenging to identify directly from experiments.
Reconstructing these fields using linear theory reveals how the full spatiotemporal structure of the instability depends on azimuthal wavenumber.
These results provide a theoretical pathway for predicting and interpreting coupled Faraday--RT dynamics in confined vibrated interfaces, particularly when complex interfacial patterns arise from the interaction of multiple modes.

\section*{Declaration of competing interest}

The authors declare that they have no known competing financial interests or personal relationships that could have appeared to influence the work reported in this paper.

\section*{Acknowledgments}

Sandia National Laboratories is a multi-mission laboratory managed and operated by National Technology \& Engineering Solutions of Sandia, LLC (NTESS), a wholly owned subsidiary of Honeywell International Inc., for the U.S. Department of Energy's National Nuclear Security Administration (DOE/NNSA) under contract DE-NA0003525. 
This written work is authored by an employee of NTESS. The employee, not NTESS, owns the right, title and interest in and to the written work and is responsible for its contents. Any subjective views or opinions that might be expressed in the written work do not necessarily represent the views of the U.S.\ Government.
The publisher acknowledges that the U.S.\ Government retains a nonexclusive, paid-up, irrevocable, world-wide license to publish or reproduce the published form of this written work or allow others to do so, for U.S. Government purposes.
The DOE will provide public access to results of federally sponsored research in accordance with the \href{https://www.energy.gov/downloads/doe-public-access-plan}{DOE Public Access Plan}.

\section*{Data availability}

Code available at \url{https://github.com/tic173/Vibrated-Interface}.



\appendix
\setcounter{figure}{0}
\setcounter{equation}{0}
\renewcommand{\theequation}{{\rm A}.\arabic{equation}}
 \renewcommand{\thefigure}{A.\arabic{figure}}

\setcounter{table}{0}
\renewcommand{\thetable}{A.\arabic{table}}

\counterwithout*{equation}{section}
\crefalias{section}{appendix}
\crefalias{subsection}{appendix}

\section*{Appendix}

\section{Linear Floquet stability analysis}
 
\subsection{Linear solution}\label{sec: Floquet_App}

We revisit the governing equation for each triplet $(m,n,i)$, where $m$ is the azimuthal wavenumber, $n$ is the Floquet harmonic index, and $i$ denotes the radial mode:
\begin{align}\label{eqn:Governing cylindrical_app}
  \left[\gamma_{m,n,i}-C^{(j)}\boldsymbol{\Delta}_m\right]\boldsymbol{\Delta}_m {\hat{w}_{m,n,i}}^{(j)}=0,
\end{align}
Here, the Floquet growth rate $\gamma_{m,n,i}$ enters the governing equation as a linear spectral parameter.
We seek separated solutions of the form for the vertical velocity component
\begin{align}
    \hat{w}_{m,n}^{(j)} (r,z)=  {\Gamma}_{m,n}(r)  {Z}_{m,n}^{(j)}(z).
\end{align}
The radial dependence is represented using normalized Bessel modes,
\begin{align}
  {\Gamma}_{m,n}(r) =  \frac{J_{|m|}(\beta_{m,n,i} r)}{J_{|m|}(\beta_{m,n,i}/\xi)},
\end{align}
where $J_{|m|}(\cdot)$ denotes the Bessel function of the first kind of order $|m|$, and
$\beta_{m,n,i}$ is the corresponding radial wavenumber.
Imposing the no-penetration condition at the lateral cylinder wall,
\begin{align}
    \left.\partial_r \hat{w}_{m,n,i}^{(j)}\right|_{r=1/\xi}=0,
\end{align}
gives the radial eigenvalue condition for the $i$th radial wavenumber
\begin{align}
    J'_{|m|}(\beta_{m,n,i}/\xi)=0. 
\end{align}
The corresponding Bessel modes, $J_{|m|}(\beta_{m,n,i} r)$, are mutually orthogonal,
\begin{align}
    \left<J_{|m|}(\beta_{m,\cdot,i_1} r), J_{|m|}(\beta_{m,\cdot,i_2} r)\right>_r = \int_{0}^{1/\xi} J_{|m|}(\beta_{m,\cdot,i_1} r), J_{|m|}(\beta_{m,\cdot,i_2} r)\, r\dd r=\mathcal{J}_{m,\cdot,i_1}\delta_{i_1,i_2},
\end{align}
 and satisfy the radial Helmholtz eigenproblem
\begin{align}
\left( \boldsymbol{\Delta}_m^{\mathrm{H}} + \beta_{m,n,i}^2 \right) J_{|m|}(\beta_{m,n,i} r) = 0,
\end{align}
indicating that $\beta$ serves as a wavenumber-like parameter characterizing the radial structure of each mode.
As the Bessel eigenfunctions $J_{|m|}(\beta_{m,\cdot,i} r)$ form a complete orthogonal basis on $r\in\left[0,1/\xi\right]$, the modified-Bessel radial structure associated with the homogeneous solution can be expanded as
\begin{align}
  \frac{I_{|m|}( r/l_{m,n,k})}{I_{|m|}( 1/(\xi l_{m,n,k}))} = \sum_i \lambda_{m,\cdot,i,k} \frac{J_{|m|}(\beta_{m,\cdot,i} r)}{J_{|m|}(\beta_{m,\cdot,i}/\xi)},
\end{align}
with projection coefficients
\begin{align}\label{eqn: lambda}
  \lambda_{m,\cdot,i,k}&=  \frac{J_{|m|}(\beta_{m,\cdot,i}/\xi)}{ I_{|m|}( 1/(\xi l_{m,n,k}))}\frac{\left<J_{|m|}(\beta_{m,\cdot,i} r),I_{|m|}(r/l_{m,n,k} )\right>_r}{ \lVert J_{|m|}(\beta_{m,\cdot,i} r)\rVert_r^2} \nonumber\\ &= \frac{2 \xi l_{m,n,k}}{1+\beta_{m,\cdot,i}^2l_{m,n,k}^2}\frac{I_{|m|}'(1/(\xi l_{m,n,k}))}{I_{|m|}(1/(\xi l_{m,n,k}))}\frac{\beta_{m,\cdot,i}^2}{\beta_{m,\cdot,i}^2-\xi^2m^2}.
\end{align}
The same form of projection remains applicable when $l_{m,n,k}$ is imaginary, in which case the homogeneous contribution is equivalently represented by a $J_{|m|}$-type radial mode.
Therefore, the total effective interface displacement component for the triplet $(m,n,i)$ is
\begin{align}
    \hat{\zeta}_{m,n,i}= -\lambda_{m,\cdot,i,k}\hat{\zeta}_{m,n, k}^{(\mathrm{H})}+\hat{\zeta}^{(\mathrm{P})}_{m,n,i}.
\end{align}

Substituting the radial dependence into the governing equations~\cref{eqn:Governing cylindrical_app}, the axial mode associated with each Floquet harmonic can be written in the general form
\begin{align}\label{eqn:w_genearal}
   Z^{(j)}_{m,n,i}(z) 
    &=\hat{\zeta}_{m,n,i} \left(a^{(j)}_{m,n,i}  \mathrm{e}^{\beta_{m,\cdot,i} z}+  b^{(j)}_{m,n,i}  \mathrm{e}^{-\beta_{m,\cdot,i}z}+c^{(j)}_{m,n,i} \mathrm{e}^{ \beta_{m,\cdot,i}q_{m,n,i}^{(j)} z}+  d^{(j)}_{m,n,i} \mathrm{e}^{- \beta_{m,\cdot,i}q_{m,n,i}^{(j)}z}\right),
\end{align}
where $q_{m,n,i}^{(j)}\equiv\sqrt{1+\gamma_{m,n,i}/(C^{(j)}\beta^2_{m,\cdot,i})}$.
The no-slip boundary conditions at the top and bottom cylinder walls require that both the velocity and its vertical derivative vanish, giving
\begin{align}\label{eqn:BC_vertical}
    \hat{w}_{m,n,i}^{(d)} = \partial_{z} \hat{w}_{m,n,i}^{(d)} = 0 \quad &\text{at} \quad z=-1, \\
    \hat{w}_{m,n,i}^{(l)} = \partial_{z} \hat{w}_{m,n,i}^{(l)} = 0 \quad &\text{at} \quad z=1.
\end{align}
At the interface $z=\zeta$, the continuity of velocity and tangential stress, and the kinematic boundary condition, yield
\begin{align}
     \left[1,\,\partial_{z},\, \eta \left(\partial_{zz}+{\beta_{m,\cdot,i}^2}\right)\right] \hat{w}_{m,n,i}^{(l)}&= \left[1,\,\partial_{z},\,  \left(\partial_{zz}+{\beta_{m,\cdot,i}^2}\right)\right] \hat{w}_{m,n,i}^{(d)},\\
   \gamma_{m,n,i} \hat{\zeta}_{m,n,i} &=\hat{w}_{m,n,i}\vert_{z=0}.\label{eqn:BC8}
\end{align}
For a given azimuthal wavenumber--mode pair $(m,\cdot,i)$, substituting the general solution in equation~\cref{eqn:w_genearal_app} into the above boundary conditions leads to an 8-equation system,
\begin{align}\label{eqn:8eqn}
     &{\vb*{Q}}
\mqty(a^{(d)}, b^{(d)},c^{(d)},d^{(d)},a^{(l)}, b^{(l)},c^{(l)},d^{(l)})^\transpose  = 
     \mqty(\gamma, 0,0,0,0,0,0,0)^\transpose.
\end{align}
Here, the dependence on $m$, $n$, and $i$ is omitted for simplicity.
The resulting coefficients determine the axial structure of the corresponding Floquet mode.

For the free-sliding configuration, the pressure-jump condition~\cref{eqn:pressure jump_coeffs} can be matched independently for each radial Bessel mode, giving
\begin{align}
& a \left(\hat{\zeta}_{m,n-1,i}+\hat{\zeta}_{m,n+1,i}\right)
- \left(2g_{\text{sgn}} +\frac{\kappa}{\mathrm{Bd}} \beta_{m,\cdot,i}^2\right) \hat{\zeta}_{m,n,i} \nonumber \\
&=
     \left[ \frac{\gamma_{m,n,i}}{\beta_{m,\cdot,i}^2}+ 3\kappa C(1-\eta)  \right]\partial_z \hat{w}_{m,n,i}^{(d)} -\frac{\kappa C}{\beta_{m,\cdot,i}^2}\left[\partial_{zzz}\hat{w}_{m,n,i}^{(d)}-\eta \partial_{zzz}\hat{w}_{m,n,i}^{(l)}\right].
\end{align}
Collecting these relations over all Floquet harmonics gives
\begin{align}
  &\left[ \underbrace{ \mqty(\ddots & \vdots &  \vdots & \vdots & \vdots \\
    \cdots  & A_{m,-1,i}& 0&0  & \cdots\\
    \cdots  & 0& A_{m,0,i}&0  & \cdots\\
    \cdots  & 0& 0&A_{m,1,i}  & \cdots\\
    \vdots & \vdots & \vdots &  \vdots &\ddots
    )}_{\vb*{A}_{m,\cdot,i}}
- a \underbrace{ \mqty(\ddots & \vdots & \vdots  & \vdots &\vdots \\
    \cdots  & 0& 1&0  & \cdots\\
    \cdots  & 1& 0&1  & \cdots\\
    \cdots  & 0& 1&0  & \cdots\\
    \vdots & \vdots & \vdots & \vdots  &\ddots
    )}_{\vb*{B}} \right]
       \underbrace{ \mqty(\vdots\\ \hat{\zeta}_{m,-1,i}\\\hat{\zeta}_{m,0,i}\\\hat{\zeta}_{m,1,i}\\\vdots)}_{\hat{\vb*{\zeta}}_{m,\cdot,i}} =0. \label{eqn:General EVD}
\end{align}
Here, $\vb*{A}_{m,\cdot,i}$ contains the individual harmonic contributions, while $\vb*B$ couples adjacent Floquet harmonics through the periodic forcing.
The diagonal coefficient for each triplet $(m,n,i)$ is
 \begin{align}
    & A_{m,n,i}(\gamma_{m,n,i};\beta_{m,\cdot,i}) \equiv \, \beta 
        \left(a^{(d)}-b^{(d)}+c^{(d)}q^{(d)}-d^{(d)}q^{(d)}\right)
        \left(\gamma/({\beta^2})+3\kappa C(1-\eta) \right) \label{eqn:An} \\
    & - \kappa C \beta 
        \left[
            a^{(d)}-b^{(d)}+c^{(d)}{q}^{(d)^3}-d^{(d)}{q}^{(d)^3} - 
            \eta
            \left(a^{(l)}-b^{(l)}+c^{(l)}{q}^{(l)^3}-d^{(l)}{q}^{(l)^3}\right)
        \right]
        + (2g_{\text{sgn}}+{\kappa\beta^2}/\mathrm{Bd}) \nonumber,
\end{align}
and the remaining coefficients $(\cdot)^{(l)}$ and $(\cdot)^{(d)}$ are obtained from equation~\cref{eqn:8eqn}.

\subsection{Validation}\label{sec:validation}

We validate the present linear Floquet stability analysis by considering two limiting cases with only RT or only Faraday instability, and by comparing the resulting growth rates with classical theory. 
For the RT limit, we consider the dimensionless dispersion relation for the infinite-depth instability~\citep{kull1991theory}, which gives
\begin{align}
 0=  & \left[\gamma^2-\left( \At k_m-\frac{1+\At}{2\mathrm{Bd}}k_m^3\right)\right]\left(\eta(1-\At) k_m+(1+\At)q_m^{(d)}+(1+\At) k_m+\eta(1-\At)q_m^{(l)}\right)\nonumber\\
     &+2\gamma k_m C\left(\eta(1-\At) k_m+(1+\At)q_m^{(d)}\right)\left((1+\At) k_m+\eta(1-\At)q_m^{(l)}\right),
\end{align}
where $q_m^{(j)}=\sqrt{k_m^2+\gamma/C^{(j)}}$, and the effective wavenumber
\begin{align}
    k_m = \beta_m
\end{align}
relates the planar dispersion relation with the discrete radial modes admitted by the cylindrical geometry by satisfying $J'_{|m|}(k_m/\xi)=0$.

\begin{figure}[ht!]
  \centering
  \includegraphics{ 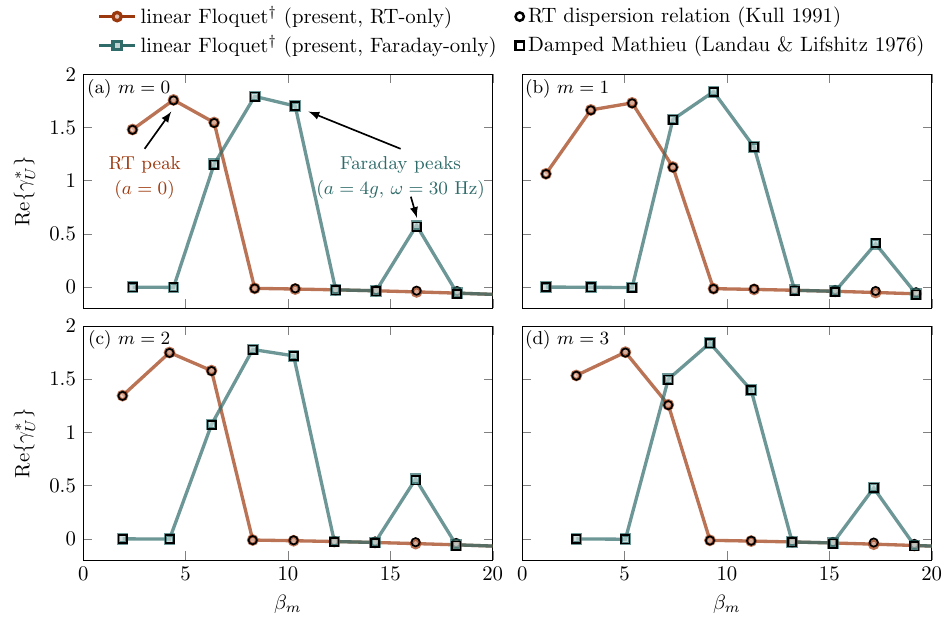}
  \caption{
    Comparison of growth rates obtained from the present linear Floquet stability analysis and classical theory for a water--air interface at different azimuthal wavenumbers:  (a) $m=0$; (b) $m=1$; (c) $m=2$; and (d) $m=3$.
    The RT-only limit corresponds to $a=0$ and $g_{\text{sgn}}=-1$, whereas the Faraday-only configuration corresponds to imposed vibration with $a=4g$ and $\omega=30~\unit{\hertz}$.
  }
  \label{fig:validation} 
\end{figure}

For the Faraday limit, we compare with the weakly viscous damped Mathieu equation for an equal-depth two-fluid interface~\citep{landau1976mechanics},
\begin{align}
       \dv{}{t} \mqty[{\zeta}_m \\ \dot{\zeta}_m] = 
       \mqty[ 0 & 1 \\
       -\tilde{\omega}_{m}^2-a \At k_m\tanh{(k_m)}\cos{(\omega t)} & -2C(1+\At)(1+\eta)k_m^2
       ] 
       \mqty[{\zeta}_m \\ \dot{\zeta}_m],
\end{align}
where $\tilde{\omega}_m\equiv \sqrt{(\At k_m+(1+\At)k_m^3/(2\mathrm{Bd}))\tanh{(k_m)}}$ is the natural frequency associated with the $m$th azimuthal wavenumber.
The monodromy matrix is obtained by integrating this linear system over one vibration period, $T=2\pi/\omega$, with the identity matrix as the initial condition. 
The corresponding Floquet growth rate is then computed as
\begin{align}
    \gamma=\frac{1}{T}\log{\left(\max_i|\mu_i|\right)},
\end{align}
where $\mu_i$ are the Floquet multipliers, obtained as the eigenvalues of the monodromy matrix.

\Cref{fig:validation} compares the growth rates predicted by the present linear Floquet stability analysis with those obtained from the corresponding classical theories. 
A favorable agreement is observed for all azimuthal wavenumbers considered, validating the present analysis in both the RT-only and Faraday-only limits.

\begin{figure}[h!]
  \centering
  \includegraphics{ 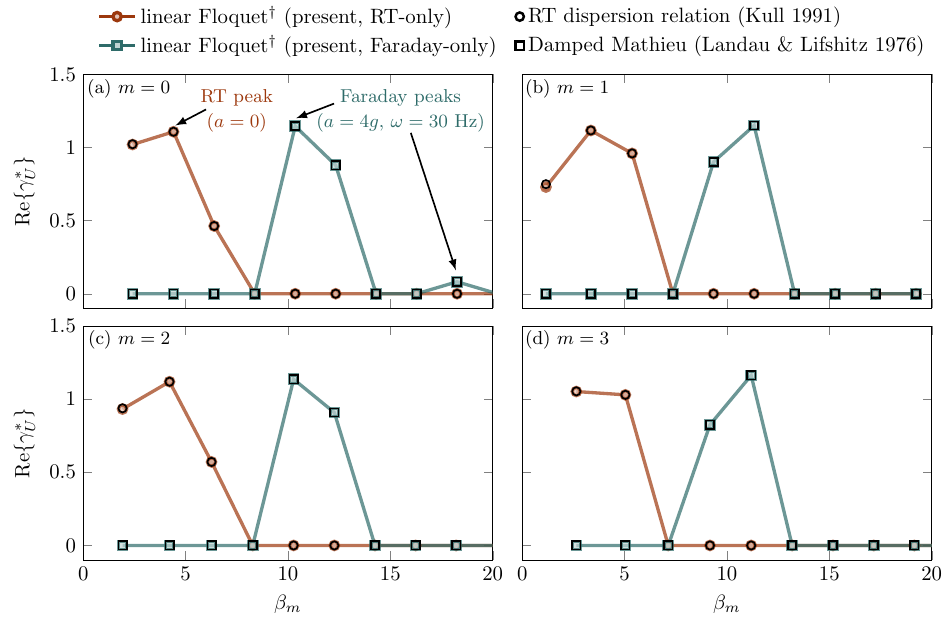}
  \caption{
    Same as \cref{fig:validation} but for a water--model-fluid interface with $\At=0.5$ and $\eta=0.3$.
  }
  \label{fig:validation_case2} 
\end{figure}

\Cref{fig:validation_case2} shows the predicted growth rates for a water--model-fluid interface with different density and viscosity ratios.
For the range of radial wavelengths considered, higher-order Faraday harmonics occur only in the axisymmetric component.
Excellent agreement is again obtained between the present analysis and the corresponding classical theories.

\begin{figure}[ht!]
  \centering
  \includegraphics{ 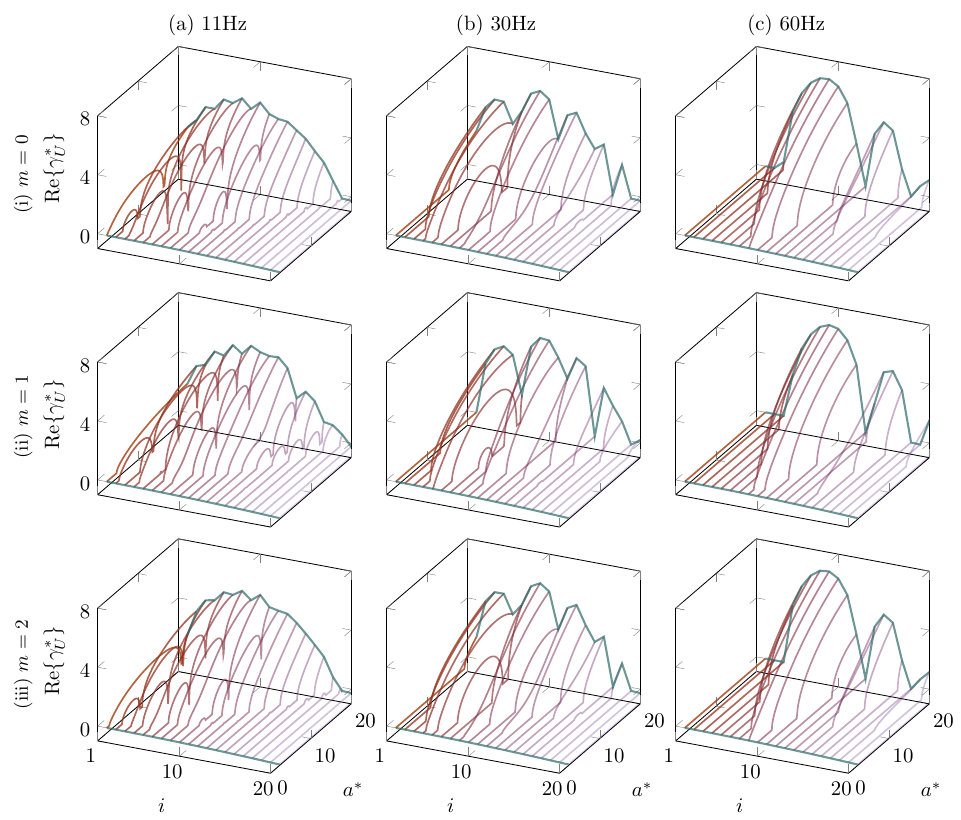}
  \caption{
    Growth rates of the axisymmetric ($m=0$; i), dipolar ($m=1$; ii) and quadrupolar ($m=2$; iii) modes for the Faraday-only case ($g_{\text{sgn}}=1$) at labeled vibration frequencies.
  }
  \label{fig:Growthrate_Faraday} 
\end{figure}

\begin{figure}[ht!]
  \centering
  \includegraphics{ 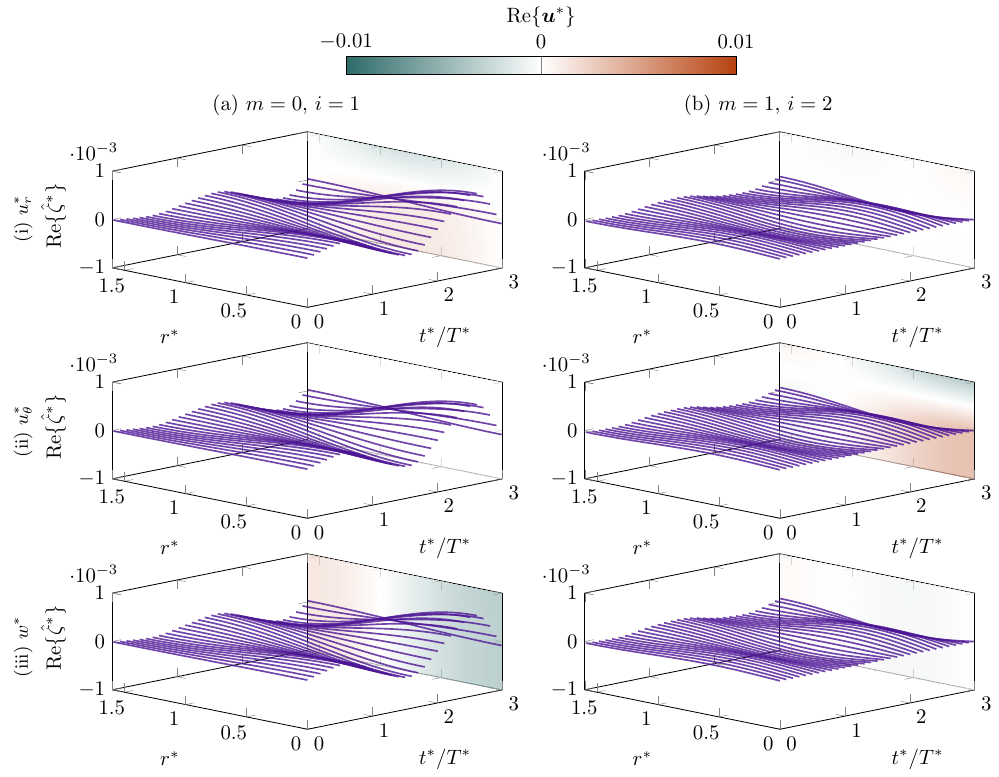}
  \caption{
    Linear-Floquet-theory predictions of the interfacial displacement and velocity fields at $t^*/T^*=3$ for the Faraday-only configuration ($g_\text{sgn}=1$) at vibration amplitude $a^*=0.9$ and frequencies as labeled.
  }
 \label{fig:Interface_Faraday}
\end{figure}

\subsection{Confinement and the long-wavelength RT cutoff}\label{sec:confinement_app}

Vibration can stabilize a confined family of RT-unstable modes, though a set of long-wavelength modes remains in an unbounded domain~\citep{chu2025competing}.
The no-penetration condition $J'_{|m|}(\beta_{m,\cdot,i}/\xi)=0$ admits radial wavenumbers $\beta_{m,\cdot,i}=\xi\,j'_{|m|,i}$, where $j'_{|m|,i}$ is the $i$th positive stationary point of $J_{|m|}$.
A mode is RT-unstable below $k_c=\sqrt{2\,\mathrm{Bd}\,\mathrm{At}/(1+\mathrm{At})}$, so azimuthal direction carries a finite count of unstable radial modes, with the longest as $\beta_{m,\cdot,1}=\xi\,j'_{|m|,1}$.
The smallest wavenumber is bounded away from zero and vanishes as $\xi\to0$.
In an unbounded domain, the spectrum is $[0,\infty)$ and the unstable modes are in $(0,k_c)$, potentially carrying arbitrarily small wavenumbers.

The gap in the stability spectrum determines the dynamic-stabilization threshold via the Mathieu reduction of \cref{sec:validation}.
For $g_{\mathrm{sgn}}=-1$ and weak damping, the mode amplitude follows $\ddot{\zeta}_\beta+(-\Omega^2+p\cos\omega t)\zeta_\beta=0$ with $\Omega^2=[\mathrm{At}\,\beta-(1+\mathrm{At})\beta^3/(2\,\mathrm{Bd})]\tanh\beta$ and $p=a\,\mathrm{At}\,\beta\tanh\beta$.
The (inverted) equilibrium is dynamically stabilized at $p^2/(2\omega^2)>\Omega^2$.
The leading-order Mathieu case is $\delta=-\epsilon^2/2$, where $\delta=-\Omega^2/\omega^2$ and $\epsilon=p/\omega^2$, and thus the acceleration exceeds
\begin{align}\label{eqn:ac}
  a_c(\beta) = \omega\left[\frac{2}{\mathrm{At}\,\beta\tanh\beta}
    \left(1-\frac{\beta^2}{k_c^{2}}\right)\right]^{1/2}.
\end{align}
This threshold decreases monotonically for increasing mode numbers.
Thus, one finite amplitude $a>a_c(\beta_{m,\cdot,1})$ stabilizes all RT-unstable modes.
In the unbounded case, this same threshold diverges as $\beta\to0$, where $a_c\sim\omega\sqrt{2/\mathrm{At}}\,/\beta$.
As such, the longest waves see no finite mode amplitudes, and a set of long-wavelength RT modes exists for all $a$.

The above considers only the RT-unstable modes, so the interface need not be globally stable.
For $a>a_c(\beta_{m,\cdot,1})$, the shorter-wavenumber radial modes can be Faraday-unstable, though suppression requires the RT- and Faraday-stable regions to overlap.
We do not observe this overlap (see \cref{fig:phase space_case2}).
The estimate of equation~\cref{eqn:ac} is the high-frequency leading-order threshold.
In the $f\to0$ limit, this is replaced by the quasi-static, capillary-limited configuration, discussed in \cref{sec:results_pinned}.
Note that viscosity does not change the long-wave observation, as the damping rate $2 \, C(1+\mathrm{At})(1+\eta)\beta^2$ vanishes faster than the RT growth rate $\sqrt{\mathrm{At}}\,\beta$ as $\beta\to0$.

\subsection{Faraday-only configuration for free-sliding interface}\label{sec:faraday_freesliding_app}

\Cref{fig:Growthrate_Faraday} shows the growth rates for the Faraday-only case ($g_{\text{sgn}}=1$) at different vibration amplitudes and frequencies.
The Faraday peaks emerge as the vibration amplitude increases.
The axisymmetric ($m=0$) and quadrupolar ($m=2$) modes show similar trends.
The dipolar mode ($m=1$), however, sees higher-order Faraday peaks at higher oscillation amplitudes, beyond the fundamental response.
Together, this predicts Faraday-type responses in bounded cylinders.

\Cref{fig:Interface_Faraday} shows the flow field and interfacial response in the Faraday-only configuration ($g_\text{sgn}=1$) at vibration amplitude $a=0.9$ and frequency $\omega=11~\unit{\hertz}$, as predicted by linear Floquet theory.
\Cref{fig:Growthrate_Faraday} shows that the axisymmetric and dipolar components are unstable in this regime.
The growth rates have similar real parts, but the interfacial dynamics are different.
Axisymmetric modes are larger-amplitude vibrations than the dipolar one, increasingly so as $r \to 0$.
Note that any non-axisymmetric azimuthal component cannot represent the displacement at the axis.
Rather, it can only be balanced by an $m=0$ contribution, which may involve additional radial modes.
The corresponding flow fields also differ.
A non-zero azimuthal velocity component accompanies the dipolar response, though the axisymmetric mode has, by definition, $u_\theta=0$.

\clearpage\newpage
\bibliography{ref.bib}

\end{document}